\documentclass[aps,pra,reprint,floatfix,amsmath,amssymb,superscriptaddress,longbibliography]{revtex4-2}
\usepackage[T1]{fontenc}
\usepackage[utf8]{inputenc}
\usepackage{bm,mathtools,amsthm}
\usepackage{tikz}
\usepackage{pgfplots}
\usetikzlibrary{arrows.meta,positioning,calc}
\usepgfplotslibrary{groupplots}
\pgfplotsset{compat=1.18}
\usepackage{xcolor}
\usepackage{hyperref}
\hypersetup{colorlinks=true,citecolor=blue,linkcolor=blue,urlcolor=blue}
\newcommand{\Tr}{\operatorname{Tr}}
\newcommand{\diag}{\operatorname{diag}}
\newcommand{\Id}{\mathbb{I}}
\newcommand{\fp}{\mathrm{fp}}

\newcommand{\mc}{C_{\mathrm{mem}}}
\newcommand{\jn}{J_N}
\newcommand{\je}{J_E}
\begin{document}
\title[Memory--transport competition in two-bath resets]{Memory--Transport Competition in Physical Two-Bath Quantum Reset Dynamics}
\author{Jishad Kumar}
\affiliation{MSP Group, Department of Applied Physics, Aalto University, P.O. Box 15600, FI-00076 Aalto, Espoo, Finland}
\author{Achilleas Lazarides}
\affiliation{Interdisciplinary Centre for Mathematical Modelling and Department of Mathematical Sciences, Loughborough University, Loughborough, Leicestershire LE11 3TU, United Kingdom}
\author{Tapio Ala-Nissila}
\affiliation{MSP Group, Department of Applied Physics, Aalto University, P.O. Box 15600, FI-00076 Aalto, Espoo, Finland}
\affiliation{Interdisciplinary Centre for Mathematical Modelling and Department of Mathematical Sciences, Loughborough University, Loughborough, Leicestershire LE11 3TU, United Kingdom}
\begin{abstract}
Partial environmental renewal must reconcile the preservation of existing quantum correlations with restoration of the reservoirs toward their prescribed thermal states. We investigate this competition in a quadratic fermionic system coupled to two independently refreshed reservoirs. A number-conserving thermal-ancilla rotation yields a completely positive channel and an exact stationary covariance equation. We establish a sharp covariance-admissibility criterion for formal hard resets, a Pauli bound on retained coherence, and an exact reciprocal mode kernel yielding a pairwise nonnegative entropy-production law. To distinguish preserving a correlated state from successfully renewing it, we derive two many-body Gaussian fidelities and bound their infidelities by changes in system--reservoir coherence. Exact finite-chain calculations reveal that a refresh can preserve the full state with high fidelity while leaving it far from its thermal product reference. A narrow, size-sensitive coherence maximum near the nonrefreshing boundary is accompanied by a substantial loss of reference fidelity. Even identical reservoirs with zero net particle current exhibit nontrivial coherence, fidelity loss, and switching work. Particle transfer supplies a complementary continuity diagnostic. These results separate state preservation, thermal renewal, and energetic irreversibility in a microscopically admissible finite-period reset protocol.
\end{abstract}
\maketitle

\section{Introduction}\label{sec:intro}
Repeated interventions modify not only the populations of an open quantum system but also the correlations through which it retains information about its environment. Misra and Sudarshan formulated the quantum Zeno effect as a limit of repeated observations \cite{MisraSudarshan1977}. Itano et al. demonstrated the suppression of transitions by repeated measurements in an atomic system \cite{Itano1990}. Kofman and Kurizki showed that frequent observations can instead accelerate decay, depending on the spectral response \cite{KofmanKurizki2000Nature}. Fischer et al. observed both regimes in an unstable quantum system \cite{Fischer2001}. Facchi and Pascazio reviewed the mathematical and physical foundations of quantum Zeno dynamics \cite{FacchiPascazio2008}. Schulman compared continuous and pulsed observations in the Zeno regime \cite{Schulman1998}. Facchi, Nakazato, and Pascazio analyzed the crossover from inhibited to accelerated decay \cite{FacchiNakazatoPascazio2001}. The dynamical formulation of Zeno subspaces was developed by Facchi and Pascazio \cite{FacchiPascazio2002}. Ruseckas and Kaulakys derived a general expression for measurement-modified transition rates \cite{Ruseckas2004}. In the present work, interventions are not projective measurements but controlled, number-conserving reservoir contacts.

Repeated-interaction models provide a microscopic description of such contacts. Scarani et al. studied thermalization and entanglement generation through repeated interactions \cite{Scarani2002}. Bruneau, Joye, and Merkli analyzed repeated and continuous interactions in open systems \cite{BruneauJoyeMerkli2010}. The collision-model review by Ciccarello et al. develops both Markovian and non-Markovian constructions \cite{Ciccarello2022PhysRep}, while Cusumano gives an introductory account of their dynamical structure \cite{Cusumano2022}. Giovannetti and Palma derived master equations for correlated quantum channels \cite{GiovannettiPalma2012}. Ciccarello, Palma, and Giovannetti constructed collision models for non-Markovian dynamics \cite{Ciccarello2013PRA}. Kretschmer et al. developed a further collision-model description of non-Markovian dynamics \cite{Kretschmer2016}. Lorenzo et al. constructed non-Markovian piecewise dynamics from collision models \cite{Lorenzo2017}. Multipartite collision schemes can also simulate general multipartite Markovian dynamics, as shown by Cattaneo et al. \cite{Cattaneo2021}. Correlations generated by collisions can be useful resources: \c{C}akmak et al. demonstrated robust multipartite entanglement generation in a collision model \cite{Cakmak2019}. Giovannetti's dynamical model of quantum memory channels illustrates why stored environmental correlations and operational information-transmission capacity must be distinguished \cite{Giovannetti2005}.

For quadratic number-conserving fermionic Hamiltonians every one-body expectation value evolves exactly through the single-particle density matrix (SPDM). Peschel showed how correlation functions determine reduced density matrices of free lattice systems \cite{Peschel2003}. Peschel and Eisler reviewed the construction of reduced density matrices and entanglement entropies in free lattice models \cite{PeschelEisler2009}. Hackl and Bianchi formulated the geometry of bosonic and fermionic Gaussian states in terms of K\"ahler structures \cite{HacklBianchi2021}. Bravyi developed a representation of fermionic linear optics and Gaussian maps \cite{Bravyi2005}. Greplov\'a and Giedke characterized degradability for fermionic Gaussian channels, including the special role of attenuation \cite{GreplovaGiedke2018}. Vieira et al. introduced an SPDM resetting framework contrasting repeated-interaction (RI) and evolving-correlations (EC) prescriptions \cite{Vieira2020}. Our subsequent study investigated coherence-controlled Zeno dynamics \cite{jishad2026}. A preceding single-reservoir investigation addressed the thermodynamics and physical admissibility of coherence-selective resetting \cite{Kumar2026CoherenceCost}. Here we ask how independently renewing two reservoirs changes the preservation of the full correlated state, the fidelity of environmental renewal, and the associated energetic and entropic exchange.

A formal rule that overwrites a bath's occupation block while independently retaining arbitrary system--bath coherences need not be a physical quantum operation. Both positivity and the fermionic Pauli upper bound constrain the allowed cross blocks, especially near almost empty or occupied bath modes. We construct a microscopic alternative in which each bath mode is mixed with a fresh thermal ancilla. The two retention amplitudes are independent between reservoirs, but each is inseparably linked to its own bath-population renewal probability. The resulting channel is completely positive on the full Fock space and closes exactly as a discrete Lyapunov equation for the covariance. We also establish a necessary and sufficient all-input covariance-admissibility threshold for the formal overwrite; it is not a complete-positivity criterion.

The two-reservoir geometry permits a richer distinction between state preservation and successful renewal. Uhlmann introduced a transition probability for mixed states \cite{Uhlmann1976}; Jozsa formulated the squared fidelity and its properties in density-operator language \cite{Jozsa1994}. We use this many-body fidelity rather than normalizing the SPDM by its trace. One fidelity compares the states immediately before and after the intervention; another compares the output with an uncorrelated thermal reference having the same system occupation. Their exact Gaussian evaluation requires only single-particle matrix operations. The bounds of Fuchs and van de Graaf relate fidelity to trace-distance distinguishability \cite{FuchsGraaf1999}; here they yield explicit constraints from retained and removed one-body coherences. Banchi, Giorda, and Zanardi used fidelity geometry to investigate dissipative criticality of Gaussian fermionic stationary states \cite{BanchiGiordaZanardi2014}. The present question is different: we compare states across a single intervention and with its thermal reference. Neither measure is an information-storage capacity. In particular, the trace-distance information-backflow criterion of Breuer, Laine, and Piilo \cite{BreuerLainePiilo2009} and the entanglement-based approach of Rivas, Huelga, and Plenio \cite{RivasHuelgaPlenio2010} concern dynamical properties that are not inferred from a stationary coherence norm.

The energetic accounting is equally important. Strasberg et al. developed a unified thermodynamic framework for repeated interactions \cite{Strasberg2017}. Strasberg extended repeated-interaction thermodynamics to strong coupling \cite{Strasberg2019}. Prositto et al. examined relaxation and energetic costs of equilibrium and nonequilibrium steady states in repeated-interaction protocols \cite{Prositto2025}. Landi and Paternostro reviewed irreversible entropy production \cite{LandiPaternostro2021}. Ptaszy\'nski and Esposito distinguished quantum and classical contributions to entropy production in Gaussian systems \cite{PtaszynskiEsposito2023}. In our setting the bath energy change alone does not quantify a universal resource cost: switching the ancilla contacts in the presence of system--bath interactions exchanges work, and the complete outgoing-ancilla state enters the entropy balance.

We use a single fermionic level coupled to two finite tight-binding chains. Their finite bandwidths permit near-edge regimes, and their known standing-wave eigenmodes allow exact finite-period propagation. Particle exchange is retained as a complementary diagnostic of the two-reservoir geometry. Relaxation-induced transport turnover is already known from extended-reservoir theory: Gruss, Velizhanin, and Zwolak identified a Kramers-like crossover \cite{Gruss2016}, and Gruss, Smolyanitsky, and Zwolak studied the corresponding relaxation-limited current regimes \cite{Gruss2017}. Our focus instead is the physical reset channel, its covariance constraints and full-state fidelity, its work and entropy ledger, and the size-sensitive structure of retained correlations. In particular, an identical-reservoir benchmark with symmetric renewal isolates these effects without net particle throughput.

Section~\ref{sec:model} defines the microscopic channel. Section~\ref{sec:exact} develops the stationary covariance, coherence constraints, and thermodynamic identities. Section~\ref{sec:short} derives the short-cycle asymptotics. Section~\ref{sec:results} presents the structured-bath numerical benchmark, beginning with full-state fidelity (including the unbiased case), followed by the coherence and energetic operating landscapes and the high-retention boundary layer. Section~\ref{sec:hard} treats the algebraic overwrite and its physicality bounds, and Sec.~\ref{sec:disc} discusses the results and limitations. The appendices give detailed derivations, fidelity evaluation, and numerical checks.

\section{Model and implementable two-bath refresh}\label{sec:model}
We set $\hbar=k_{\mathrm B}=1$. The number-conserving fermionic Hamiltonian is
\begin{equation}\hat H=\sum_{\alpha\beta}\hat a_\alpha^\dagger M_{\alpha\beta}\hat a_\beta,\qquad M=M^\dagger.\label{eq:H}\end{equation}
For a single system mode $|0\rangle$ coupled to $N_{\mathrm L}$ and $N_{\mathrm R}$ bath modes,
\begin{equation}
M=\begin{pmatrix}\omega_0&g_{\mathrm L}&g_{\mathrm R}\\g_{\mathrm L}^\dagger&M_{\mathrm L}&0\\g_{\mathrm R}^\dagger&0&M_{\mathrm R}\end{pmatrix},\quad
M_a=\diag(\omega_{ak}),\label{eq:M}
\end{equation}
where $g_a=(g_{a1},\ldots,g_{aN_a})$ is a row vector; the full one-particle Hilbert space is $S\oplus E_{\mathrm L}\oplus E_{\mathrm R}$. There is no direct left--right Hamiltonian hopping: the reservoirs exchange particles only through the distinguished system level. The bath energies $\omega_{ak}$ and couplings $g_{ak}$ are allowed to be arbitrary at this stage. Phases of the couplings in Eq.~\eqref{eq:M} can be removed by independent mode rephasings, so the Hamiltonian can be made real symmetric without changing the thermal targets. The canonical fermionic operators obey $\{\hat a_\alpha,\hat a_\beta^\dagger\}=\delta_{\alpha\beta}$ and $\{\hat a_\alpha,\hat a_\beta\}=0$. Defining the annihilation column $\hat{\boldsymbol a}=(\hat a_0,\hat a_{{\rm L}1},\ldots,\hat a_{{\rm R}N_{\rm R}})^T$, the exact Heisenberg equation reads $i\partial_t\hat{\boldsymbol a}=M\hat{\boldsymbol a}$, so $\hat{\boldsymbol a}(t)=e^{-iMt}\hat{\boldsymbol a}(0)$. Consequently, no approximation is needed to propagate the following one-body quantity.

The SPDM is $C_{\alpha\beta}=\langle\hat a_\beta^\dagger\hat a_\alpha\rangle$. Its fermionic admissibility criterion is $0\le C\le\Id$; for quadratic number-conserving dynamics its exact unitary update is
\begin{equation}C_n^-=U C_n^+U^\dagger,\qquad U=e^{-iM\tau}.\label{eq:U}\end{equation}
Let $t_n=n\tau$. We write $C_n^+=C(t_n^+)$ for the state immediately \emph{after} the $n$th refresh and $C_n^-=C(t_{n+1}^-)$ for the state immediately \emph{before} the next refresh; hence one cycle is first the exact unitary update $C_n^-=U C_n^+U^\dagger$ and then the ancilla contact $C_{n+1}^+=\mathcal R(C_n^-)$. The time superscripts refer to the two sides of the cycle boundary, not to positive or negative energies. A cycle-stationary state obeys $C_{n+1}^+=C_n^+$ and may differ from its pre-refresh counterpart $C_n^-$. The thermal target covariances are
\begin{equation}F_a=\diag(f_{ak}),\qquad f_{ak}=\frac{1}{e^{\beta_a(\omega_{ak}-\mu_a)}+1}.\label{eq:F}\end{equation}

\subsection{Microscopic refresh channel}
A complete reset of a bath mode is recovered by swapping it with a prepared ancilla, but a continuously variable \emph{physical} refresh must specify the joint operation producing the retained correlations. We choose the simplest such operation: passive mixing. Introduce a fresh, uncorrelated fermionic mode for every reservoir eigenmode $\hat b_{ak}$, initially with occupation $f_{ak}$. A number-conserving two-mode rotation acts, with real $0\le\eta_a\le1$, as
\begin{align}
\hat a_{ak}^+&=\eta_a\hat a_{ak}^-+\sqrt{1-\eta_a^2}\,\hat b_{ak},\label{eq:rotate}\\
\hat b_{ak}^{\mathrm{out}}&=-\sqrt{1-\eta_a^2}\,\hat a_{ak}^-+\eta_a\hat b_{ak}.\nonumber
\end{align}
Each two-mode transformation is an orthogonal rotation in one-particle mode space, and is therefore induced by a number-conserving unitary on the two-mode Fock space. Fresh ancillas have no correlations with the previously retained system and baths. No measurement or postselection is involved. Tracing out the outgoing ancillas defines a fermionic Gaussian channel. Setting
\begin{align}
K&=\diag(1,\eta_{\mathrm L}\Id_{N_{\mathrm L}},\eta_{\mathrm R}\Id_{N_{\mathrm R}}),\nonumber\\
D&=\diag\bigl(0,(1-\eta_{\mathrm L}^2)F_{\mathrm L},(1-\eta_{\mathrm R}^2)F_{\mathrm R}\bigr).\label{eq:KD}
\end{align}
the exact post-refresh covariance is
\begin{equation}C_{n+1}^+=K C_n^-K+D.\label{eq:refresh}\end{equation}
Thus the system occupation is unchanged, each system--bath block is scaled by $\eta_a$, the bath occupation matrices relax \emph{partially} toward their targets, and the cross-bath block is scaled by $\eta_{\mathrm L}\eta_{\mathrm R}$. The attenuation parameter changes both coherence retention and the refresh strength $1-\eta_a^2$: these are linked controls, not independently selectable thermodynamic interventions. In particular,
\begin{align}
C_{aa,n+1}^+&=\eta_a^2C_{aa,n}^-+(1-\eta_a^2)F_a,\label{eq:bathmap}\\
C_{0a,n+1}^+&=\eta_a C_{0a,n}^-,\qquad
C_{\mathrm L\mathrm R,n+1}^+=\eta_{\mathrm L}\eta_{\mathrm R}C_{\mathrm L\mathrm R,n}^-.\nonumber
\end{align}
The parameter extremes have different microscopic meanings. At $\eta_a=0$ the incoming fresh ancilla replaces every bath eigenmode on side $a$, while the pre-contact bath mode exits through the outgoing ancilla port. At $\eta_a=1$ no bath mode on that side is renewed or attenuated. Intermediate values both damp existing cross-correlations and partially restore occupations. Full bath replacement is recovered at $\eta_a=0$, but then that side's system--bath coherence must also vanish. Unlike a hard replacement with retained cross correlations, Eq.~\eqref{eq:refresh} has an explicit quantum-channel realization. In the tight-binding application, the attenuators act independently on \emph{bath eigenmodes}, which are extended in real space. The construction is side-local (left and right bath operations are independent), but it is not automatically implementable by contact with just the endpoint site of a spatially local chain. Alternatively, identical site-to-site rotations between a retained chain and an identical thermal auxiliary chain realize the same channel when the attenuation is uniform over the side: the common transformation to standing-wave modes commutes with these rotations. Contact only at the endpoint would not realize this channel.

\section{Exact stationary solution and observables}\label{sec:exact}
\subsection{Discrete Lyapunov equation and physical stationary state}
Define $A=KU$ and combine Eqs.~\eqref{eq:U} and \eqref{eq:refresh}:
\begin{equation}C_{n+1}^+=A C_n^+ A^\dagger+D.\label{eq:affine}\end{equation}
Whenever $r(A)<1$, the unique stationary covariance satisfies the discrete Lyapunov equation
\begin{equation}C_\fp=A C_\fp A^\dagger+D,\label{eq:stein}\end{equation}
with absolutely convergent solution
\begin{equation}C_\fp=\sum_{m=0}^\infty A^mD(A^\dagger)^m.\label{eq:series}\end{equation}
Equivalently, using column-wise vectorization,
\begin{equation}\operatorname{vec}C_\fp=[\Id-(A^*\otimes A)]^{-1}\operatorname{vec}D.\label{eq:vec}\end{equation}
No Markov, secular, or Born approximation enters this finite-size result. The attenuation matrix obeys $\lVert K\rVert\le 1$ and $U$ is unitary. In finite dimension the condition $r(A)<1$ is equivalent to the absence of an eigenvector of $U$ that lies entirely in the undamped subspace $\mathcal H_1=\ker(\Id-K^2)$. For both $\eta_a<1$ this subspace contains only the system level, whereas if $\eta_{\rm L}=1$ and $\eta_{\rm R}<1$ it includes the system and \emph{all} left-bath modes. Therefore a one-sided nonrefreshing boundary requires the stronger, full-subspace spectral check; the absence of attenuation on one side alone does not prove either existence or nonexistence of an attractive fixed point. We give the necessity and sufficiency proof in Appendix~\ref{app:fullblocks}. The no-refresh corner $\eta_{\rm L}=\eta_{\rm R}=1$ is excluded from the stationary optimization because unitary evolution alone does not select a unique attractive fixed point. A stationary point on a singly unrefreshed edge, if it exists, still obeys the exact terminal-blockade theorem.

Crucially, for $0\le C_n^-\le\Id$,
\begin{align}
C_{n+1}^+&=KC_n^-K+D\ge0,\nonumber\\
\Id-C_{n+1}^+&=K(\Id-C_n^-)K+(\Id-K^2)-D\ge0.\label{eq:positivity}
\end{align}
Both inequalities hold because $0\le F_a\le\Id$. Every stationary covariance obtained from this stable physical channel is admissible and determines a gauge-invariant fermionic Gaussian state. The covariance propagation itself does not require Gaussian initial states. If $r(A)<1$, the covariance attractor is unique for every admissible initial covariance. For the fidelity and entropy calculations we use the gauge-invariant Gaussian realization, which is stationary as a full density operator because the channel preserves Gaussianity and this Gaussian state is uniquely specified by its covariance. At finite inverse temperatures and finite mode energies, the stationary covariance is strictly interior, as shown in Appendix~\ref{app:cp}.

\subsection{Stationary memory, particle balance, and energy transfer}
At fixed point let $C_\fp^-=UC_\fp U^\dagger$ and set $x_a=(C_\fp)_{0a}$. We characterize retained one-body coherence by
\begin{equation}\mc=\sum_{a=\mathrm L,\mathrm R}\Tr(x_a x_a^\dagger)=\sum_{a,k}|x_{ak}|^2.\label{eq:memory}\end{equation}
This norm quantifies stored system--reservoir one-body coherence, not a channel capacity or a standard non-Markovianity monotone. It is constrained by an exact fermionic bound. Since $0\le C_\fp\le\Id$, functional calculus implies $C_\fp^2\le C_\fp$. Taking the system-level diagonal entry yields
\begin{align}
0&\le (C_\fp-C_\fp^2)_{00}
=P_\fp(1-P_\fp)-C_{\rm mem},\nonumber\\
0&\le C_{\rm mem}\le P_\fp(1-P_\fp)\le\tfrac14.
\label{eq:memorybound}
\end{align}
Thus retaining coherence is necessarily limited by the occupation and Pauli exclusion, independently of reservoir spectra, reset parameters, or cycle duration. The bound is necessary but not a complete characterization of which coherence spectra can be realized. The particles transferred to the outgoing fresh-ancilla stream per cycle are
\begin{align}
\Delta N_a&=\Tr(C_{aa,\fp}^- - C_{aa,\fp})\nonumber\\
&=(1-\eta_a^2)\Tr(C_{aa,\fp}^- - F_a),\label{eq:dN}
\end{align}
where $C_{aa,\fp}$ is the \emph{post-refresh} bath block, unlike a hard reset that fixes it to $F_a$. The continuity relation does not require assumptions about reservoir sizes or spectral profiles. Indeed, unitarity gives $\Tr C_\fp^-=\Tr C_\fp$, and the contact leaves $P$ unchanged. Subtracting the two traces therefore gives $\sum_a\Tr(C_{aa,\fp}^--C_{aa,\fp})=0$. Unitarity and stationarity imply the exact particle continuity law
\begin{equation}\Delta N_{\mathrm L}+\Delta N_{\mathrm R}=0,\qquad
\jn=\frac{\Delta N_{\mathrm R}}{\tau}=-\frac{\Delta N_{\mathrm L}}{\tau}.\label{eq:current}\end{equation}
\paragraph{Exact terminal-blockade lemma.} At any cycle-stationary state, Eq.~\eqref{eq:dN} directly gives $\Delta N_a=0$ if $\eta_a=1$. The continuity law in Eq.~\eqref{eq:current} then implies $\Delta N_b=0$ for the other terminal, and hence $J_N=0$. This is an exact, model-independent endpoint statement for this number-conserving refresh operation, valid at arbitrary $\tau$ and finite bath size wherever a stationary state exists. In contrast, $\eta_a=0$ generally permits nonzero transport because particles are exchanged during the preceding unitary leg. The lemma guarantees blockade on the full-retention edges of the closed retention square, but not an interior maximum relative to its zero-retention edges.\par
The exact finite-reservoir particle-exchange bounds, which follow directly from the Pauli constraint, are given in Appendix~\ref{app:energy}. They vanish when either reservoir is unrefreshed. 

The transfer $\Delta N_a$ is the net number gained by the \emph{outgoing} ancilla modes relative to their incoming occupations over one period, equivalently the reduction of the retained bath's population during its contact with the fresh ancillas. It is not the change of the bath from its fixed thermal target: at partial refresh the post-contact bath block need not equal $F_a$. In particular, the exact microscopic identity makes $\Delta N_a$ proportional to the refresh probability $1-\eta_a^2$ but leaves its prefactor state-dependent.

The signed energy delivered to the fresh ancillas, whose mode energies match those of the reservoir modes, is
\begin{equation}Q_a=\Tr[M_a(C_{aa,\fp}^- - C_{aa,\fp})],\quad \je=\frac{Q_{\mathrm L}+Q_{\mathrm R}}{\tau}.\label{eq:energy}\end{equation}
The $Q_a$ are \emph{signed energy transfers}, not automatically positive heats or a monotone measure of thermodynamic cost. A maximum of $J_E$ is thus a maximum of the \emph{signed} ancilla energy flux, not an optimization of minimum dissipated work or efficiency. The associated grand-canonical energy transfer is $Q_a-\mu_a\Delta N_a$; the first-law accounting of measurement processes is discussed by Jacobs \cite{Jacobs2012}. Barra identified the work required to maintain repeated boundary contacts \cite{Barra2015}. De Chiara et al. demonstrated the importance of collision-maintenance work in the thermodynamics of local master equations \cite{DeChiara2018}. Switching on the ancilla couplers when the system remains coupled to the baths can require control work, which is accounted for explicitly in Appendix~\ref{app:energy}. In particular, the bias-dependent energy and particle fluxes must not be conflated.

\subsection{Reciprocal mode kernel and entropy production}\label{sec:kernel}
Index the $N_{\rm L}+N_{\rm R}$ ancilla modes by $i=(a,k)$, with occupation $f_i$, energy $\omega_i$, and inverse-temperature-weighted grand energy $h_i=\beta_a(\omega_i-\mu_a)$. Write $F=\operatorname{diag}(f_i)$ and $H_\eta=\operatorname{diag}(\eta_a)$ on the ancilla space. Define the rectangular injection matrix $L$ by $L_{0i}=0$ and $L_{ji}=\delta_{ji}\sqrt{1-\eta_a^2}$ on the bath rows. Thus $LL^\dagger=I-K^2$ and $D=LFL^\dagger$. For $r(A)<1$, the outgoing ancilla occupation of mode $i$ is exactly $f_i^{\rm out}=\sum_j \mathcal T_{ij} f_j$, with
\begin{align}
\mathcal T_{ij}&=\eta_{a(i)}^2\delta_{ij}+\sum_{m=0}^\infty |(V_m)_{ij}|^2,\label{eq:transferkernel}\\
V_m&=L^\dagger U(KU)^m L.\nonumber
\end{align}
Here $m$ counts earlier injections; independent incoming ancilla streams contribute incoherently, while each $V_m$ retains the coherent propagation amplitudes within and between cycles. The kernel is nonnegative and doubly stochastic. For a real symmetric $M$ and real diagonal attenuations, $V_m^T=V_m$, so $\mathcal T_{ij}=\mathcal T_{ji}$. The proof, including normalization, is given in Appendix~\ref{app:kernel}. In particular, the star Hamiltonian (\ref{eq:M}) satisfies the reciprocity hypothesis after mode rephasing.

The net transfers in Eqs.~\eqref{eq:dN} and \eqref{eq:energy} are consequently governed by the exact mode imbalance
\begin{equation}
\delta n_i=f_i^{\rm out}-f_i=\sum_j\mathcal T_{ij}(f_j-f_i).
\label{eq:modeimbalance}
\end{equation}
For a symmetric kernel the within-terminal particle exchanges cancel pairwise, leaving
\begin{equation}
\tau J_N=\sum_{i\in {\rm R},j\in {\rm L}}\mathcal T_{ij}(f_j-f_i).
\label{eq:kernelcurrent}
\end{equation}
This is a discrete-cycle mode-transfer representation, not an elastic Landauer formula: the external controller permits transfer between unequal bare energies. The total entropy production per period has the pairwise form
\begin{align}
\Sigma&=\sum_i h_i\delta n_i\nonumber\\
&=\tfrac12\sum_{ij}\mathcal T_{ij}(h_i-h_j)(f_j-f_i)\ge0,
\label{eq:kernelentropy}
\end{align}
since $f_i=(1+e^{h_i})^{-1}$ decreases with $h_i$. The full information-theoretic identity holds even without reciprocity (Appendix~\ref{app:energy}). Here and in the entropy balance we assume finite inverse temperatures and finite chemical potentials; zero-temperature limits of the grand-energy expression require separate care. If both targets have identical positive temperature and chemical potential, continuity gives $\Sigma=\beta\tau J_E$, so the signed switching work is nonnegative. At unequal chemical potentials the nonnegative combination is instead Eq.~\eqref{eq:biasSecondLaw}. These statements follow from the microscopic channel rather than from an assumed cost function.

\subsection{Full-state preservation and thermal-reference fidelity}\label{sec:fidtheory}
A covariance norm does not determine the distinguishability of two full many-body states. Restricting to the gauge-invariant Gaussian stationary realization, let $\hat\rho(C)$ denote the Gaussian density operator with SPDM $C$. We adopt the squared Uhlmann fidelity
\begin{equation}
\mathcal F(C,B)=\left[\Tr\sqrt{\sqrt{\hat\rho(C)}\hat\rho(B)\sqrt{\hat\rho(C)}}\right]^2.
\label{eq:fiddef}
\end{equation}
The trace in Eq.~\eqref{eq:fiddef} is over Fock space. In particular, $C/\Tr C$ is not substituted for $\hat\rho(C)$. Define the covariance of the uncorrelated thermal reference at the same stationary system occupation,
\begin{equation}
C_{\rm ref}=\diag(P_\fp,F_{\rm L},F_{\rm R}).
\end{equation}
Two fidelities probe the intervention:
\begin{align}
\mathcal F_{\rm pres}&=\mathcal F(C_\fp^-,C_\fp),\label{eq:fpres}\\
\mathcal F_{\rm ref}&=\mathcal F(C_\fp,C_{\rm ref}).\label{eq:fref}
\end{align}
The former measures preservation across the refresh stroke, not across the entire stationary cycle (for which the fidelity is trivially unity). The latter measures the joint output's proximity to the ideal uncorrelated reset target. It is sensitive to population deviations and inter-bath as well as system--bath correlations. The reduced single-level state of the gauge-invariant Gaussian realization is unchanged across the refresh, so its own fidelity is identically one and cannot resolve the intervention.

For strictly interior covariances $0<C<I$ and $0<B<I$ (equivalently, full-rank gauge-invariant Gaussian density operators), let $R_C=C(I-C)^{-1}$ and $R_B=B(I-B)^{-1}$. An exact determinant formula, derived in Appendix~\ref{app:fidelity}, is
\begin{align}
\mathcal F(C,B)&=\det(I-C)\det(I-B)\nonumber\\
&\quad\times\left[\det\left(I+\sqrt{R_C^{1/2}R_B R_C^{1/2}}\right)\right]^2.
\label{eq:fidexact}
\end{align}
All determinants and matrix roots act in the one-particle space; singular endpoints follow by continuity. This formula retains noncommutativity of the covariances.

The reference is invariant under the refresh channel, and quantum fidelity is nondecreasing under a common completely positive trace-preserving map. Therefore
\begin{equation}
\mathcal F(C_\fp,C_{\rm ref})\ge\mathcal F(C_\fp^-,C_{\rm ref}).
\label{eq:fidcontract}
\end{equation}
This is a within-stroke inequality, not a monotonicity theorem in $\eta_a$, because changing retention also changes the stationary state. Moreover, the Fuchs--van de Graaf upper bound on trace distance gives the exact coherence constraints
\begin{align}
1-\mathcal F_{\rm pres}&\ge\sum_{a,k}(1-\eta_a)^2|(C_\fp^-)_{0,ak}|^2,\label{eq:fidpresbound}\\
1-\mathcal F_{\rm ref}&\ge C_{\rm mem}.\label{eq:fidrefbound}
\end{align}
For a proof, choose a normalized bath superposition $b$ aligned with the difference of the system--bath coherence rows of the two covariances. The Hermitian number-conserving observable $a_0^\dagger b+b^\dagger a_0$ has operator norm one, since its nonzero eigenvalues on the two-mode Fock space are $\pm1$. Its expectation difference equals twice the Euclidean norm of the row difference (the phase of $b$ is chosen accordingly). Trace-norm duality then bounds that row norm by the trace distance, which is at most $\sqrt{1-\mathcal F}$. Inserting the refresh law or the zero cross block of the reference proves the respective inequalities. These bounds are necessary, not generally saturated. At complete replacement of both baths, $C_\fp=C_{\rm ref}$ and $\mathcal F_{\rm ref}=1$ exactly. At intermediate retention, a large preservation fidelity alone does not certify faithful thermal renewal.

\section{Short-cycle coherence and exchange laws}\label{sec:short}
For any fixed $0\le\eta_{\mathrm L},\eta_{\mathrm R}<1$, as $\tau\to0$ the stationary bath covariance approaches $F_a$ up to $O(\tau^2)$ and the system--bath coherences are $O(\tau)$. The distinction between an arbitrary initial covariance and a stationary covariance matters. Writing $B_{bq,ak,n}=(C_n^+)_{bq,ak}$, the \emph{general} first-order update, for any physical transient covariance, is
\begin{align}
x_{ak,n+1}=&\eta_a\Bigl\{x_{ak,n}
-i\tau\bigl[(\omega_0-\omega_{ak})x_{ak,n}\nonumber\\
&+\sum_{bq}g_{bq}B_{bq,ak,n}-P_ng_{ak}\bigr]\Bigr\}
+O(\tau^2),\label{eq:xgeneral}
\end{align}
where $P_n=(C_n^+)_{00}$. It follows from the exact Hamiltonian commutator and retains both within-bath and inter-bath coherences. Only on the fixed-point (or a short-cycle relaxed slow) manifold, where $B_{bq,ak}=\delta_{ab}\delta_{qk}f_{ak}+O(\tau^2)$, does it reduce to
\begin{align}
x_{ak,n+1}=&\eta_a\{x_{ak,n}-i(\omega_0-\omega_{ak})x_{ak,n}\tau\nonumber\\
&\hspace{15mm}+i g_{ak}(P_n-f_{ak})\tau\}+O(\tau^2).
\label{eq:xupdate}
\end{align}
At the fixed point,
\begin{equation}x_{ak,\fp}=\frac{i\eta_a g_{ak}(P_\fp-f_{ak})}{1-\eta_a}\,\tau+O(\tau^2).\label{eq:xasym}\end{equation}
At the stationary point, or on the short-cycle slow manifold after fast coherence transients have relaxed, $x_{ak}$ has the slaved form in Eq.~\eqref{eq:xasym} (with $P_\fp$ replaced by the slowly varying $P$ in the latter case). Here $P=(C^+)_{00}$ and $P^-=(UC^+U^\dagger)_{00}$. The second-order occupation change on such a state (\emph{not} on an arbitrary initial transient covariance) follows by expanding the unitary to \emph{second} order. Including the direct intra-cycle transition probability gives
\begin{equation}
P^- -P=-\tau^2\sum_{a,k}|g_{ak}|^2\frac{1+\eta_a}{1-\eta_a}(P-f_{ak})+O(\tau^3).\label{eq:deltaP}
\end{equation}
Consequently, defining
\begin{equation}\gamma_{ak}(\eta_a,\tau)=|g_{ak}|^2\tau^2\frac{1+\eta_a}{1-\eta_a},\label{eq:gamma}\end{equation}
we obtain, with fixed bath dimension and $\eta_a$ strictly below unity,
\begin{align}
\sum_{a,k}\gamma_{ak}(P_\fp-f_{ak})&=O(\tau^3),\label{eq:Pcondition}\\
\Delta N_a&=\sum_k\gamma_{ak}(P_\fp-f_{ak})+O(\tau^3),\label{eq:currentasym}\\
Q_a&=\sum_k\omega_{ak}\gamma_{ak}(P_\fp-f_{ak})+O(\tau^3).\label{eq:Qasym}
\end{align}
When $\sum_{a,k}\gamma_{ak}>0$, the leading occupation is
\begin{equation}P_\fp=\frac{\sum_{a,k}\gamma_{ak}f_{ak}}{\sum_{a,k}\gamma_{ak}}+O(\tau).\label{eq:Pweighted}\end{equation}
The explicit short-cycle through-current obtained by eliminating the leading stationary occupation is given in Appendix~\ref{app:small}. Here the important point is that the same occupation mismatch determines the generated cross correlations and the energy deposited in the ancillas.

At \emph{complete coherence erasure}, the leading coefficient is
\begin{equation}\gamma_{ak}(0,\tau)=|g_{ak}|^2\tau^2.\label{eq:eta0}\end{equation}
For a nonzero coupling the coefficient increases as $\eta_a$ increases below unity; the apparent divergence near unity signals a \emph{nonuniform} asymptotic limit, not unbounded physical transport. The location of an interior optimum at finite $\tau$ is therefore a property of the exact two-bath operating diagram, not a theorem derived from endpoint-vanishing kernels. Appendix~\ref{app:small} proves Eqs.~\eqref{eq:deltaP}--\eqref{eq:Qasym} term by term.

\subsection{Short-cycle memory scale and limitations}
At fixed $\eta_a<1$ and fixed bath dimensions, the same coherence expansion gives
\begin{equation}
C_{\rm mem}=\tau^2\sum_{a,k}
\frac{\eta_a^2|g_{ak}|^2}{(1-\eta_a)^2}
(P_\fp-f_{ak})^2+O(\tau^3).
\label{eq:memshort}
\end{equation}
This expression is pointwise in retention parameters: the stationary occupation $P_\fp$ itself changes when either $\eta_a$ is varied. It also obeys the exact all-$\tau$ upper bound in Eq.~\eqref{eq:memorybound}; the diverging formal prefactor as $\eta_a\to1$ signals loss of asymptotic uniformity rather than a physical divergence. Neither monotonicity of the full memory function nor universal separation of memory and current optima follows from the explicit prefactor alone. In fact, at the fixed cycle duration used below both the fermionic bounds and the full finite-time coherent evolution are essential. The formal divergences in Eqs.~\eqref{eq:gamma} and \eqref{eq:memshort} when $\eta_a\to1$ do not describe physical divergences, because the short-cycle expansion loses uniformity near that endpoint. The exact terminal-blockade theorem, rather than that singular expansion, must be used on the full-retention boundary.

\section{Structured-bath numerical benchmark}\label{sec:results}
For each bath choose a nearest-neighbor semi-infinite tight-binding chain with hopping $J_a$ and a contact hopping $t_a$ between the system and its endpoint. For real $t_a$ and positive $J_a$, its eigenmode energies, continuum coupling, and spectral density are
\begin{align}
\omega_a(k)&=-2J_a\cos k,\quad g_a(k)=t_a\sqrt{\frac2\pi}\sin k,\nonumber\\
\mathcal J_a(\omega)&=\frac{t_a^2}{2\pi J_a^2}\sqrt{4J_a^2-\omega^2}\,\Theta(2J_a-|\omega|).\label{eq:spec}
\end{align}
The tight-binding description follows standard solid-state methods \cite{Grosso2014}; Green-function methods provide complementary spectral tools \cite{Economou2006}.
We use $N_a$-site open-chain approximations with $k_j=j\pi/(N_a+1)$ and $g_{aj}=t_a\sqrt{2/(N_a+1)}\sin k_j$. Unless explicitly stated, $N_{\mathrm L}=N_{\mathrm R}=28$ and
\begin{align}
J_{\mathrm L}=J_{\mathrm R}&=1,\quad t_{\mathrm L}=t_{\mathrm R}=0.25,\quad\omega_0=1.9,\nonumber\\
\tau&=0.25,\quad\beta_{\mathrm L}=\beta_{\mathrm R}=4,\nonumber\\
\mu_{\mathrm L}&=0.8,\quad\mu_{\mathrm R}=-0.8.\label{eq:params}
\end{align}
The finite-chain energies are $\omega_{aj}=-2J_a\cos[j\pi/(N_a+1)]$ and the discrete contact couplings are $g_{aj}=t_a\sqrt{2/(N_a+1)}\sin[j\pi/(N_a+1)]$. In constructing the propagator one exponentiates the complete Hermitian matrix $M$, including all inter-bath excursions mediated by the system level. This procedure retains correlations not resolved by a broad-band reservoir description. At every retention point the complete covariance is found from a matrix equation, rather than from fitting relaxation rates or iterating to an assumed convergence time.

\subsection{Fidelity, retained correlations, and thermal renewal}\label{sec:fidelityresults}
Figure~\ref{fig:fidelity} compares the two full-state fidelities at the same stationary points rather than comparing different stationary states as if they were inputs to one common channel. Along the biased symmetric cut, increasing retention from zero to $0.995$ raises the preservation fidelity from $0.997258$ to $0.999996$, while the thermal-reference fidelity falls from unity to $0.942751$. At $\eta=0.9995$ these values are $0.9999987$ and $0.109689$, respectively. A nearly nondisturbing refresh can thus leave the joint state far from the intended product reference. Fidelity is a many-body quantity; the small reference value may reflect accumulated deviations over many modes and cannot be inferred from $C_{\rm mem}$ alone.

The near-boundary cut is particularly revealing. At $(\eta_{\rm L},\eta_{\rm R})=(0.42529,0.9984)$, close to the $N_a=28$ coherence maximum, $C_{\rm mem}=0.0444952$ and $\mathcal F_{\rm pres}=0.998809$, but $\mathcal F_{\rm ref}=0.527087$. At the singly unrefreshed endpoint $(0.42529,1)$, the coherence drops to $0.0003085$ and the reference fidelity is approximately $1.76\times10^{-8}$, despite $\mathcal F_{\rm pres}=0.999157$. The low reference fidelity reflects the inability to impose the right bath's prescribed target when its contact is absent. This endpoint is not the doubly unrefreshed corner: the left side continues to renew. Neither the coherence maximum nor high state-preservation fidelity implies proximity to the thermal reference. Conversely, the small coherence at the boundary does not imply successful renewal. At the coherence maxima along the same cut for $N_a=14,28,48,64$, the reference fidelities are approximately $0.70324$, $0.52849$, $0.42465$, and $0.35832$, respectively; all preservation fidelities remain above $0.9987$. The size dependence involves both the shift in the maximizing retention and the increasing number of modes in the fidelity. These are finite-size, parameter-specific values; the fidelities are not intensive quantities and no infinite-reservoir fidelity limit is inferred.

To separate the fidelity effects from particle throughput, we additionally take $\mu_{\rm L}=\mu_{\rm R}=0$ and $\eta_{\rm L}=\eta_{\rm R}=\eta$, keeping all other parameters in Eq.~\eqref{eq:params}. Interchanging the identical chains leaves $M$, $K$, and $D$ invariant. Uniqueness of the stationary covariance therefore implies $\Delta N_{\rm L}=\Delta N_{\rm R}$; Eq.~\eqref{eq:current} gives both transfers identically zero. The stationary state is nevertheless generally not the uncorrelated thermal reference. At $\eta=0.9$ we find $C_{\rm mem}=0.00526377$, $\mathcal F_{\rm pres}=0.999830$, $\mathcal F_{\rm ref}=0.988270$, and $J_E=0.0135872$. Since both contacts have identical temperature and chemical potential, Eq.~\eqref{eq:biasSecondLaw} yields $\Sigma=\beta\tau J_E>0$. Coherence, imperfect restoration and dissipative switching work can therefore coexist in the absence of net particle transfer. At $\eta=0.9995$ the reference fidelity remains $0.982806$ for identical targets, unlike the much smaller value under chemical bias. This comparison does not assert that symmetric retention always maximizes fidelity or that all biases cause the same changes. It isolates the effects of the two reset objectives in a controlled benchmark.

\begin{figure*}[t]
\centering
\begin{tikzpicture}
\begin{groupplot}[group style={group size=3 by 1,horizontal sep=1.50cm},width=0.230\textwidth,height=0.24\textwidth,scale only axis,grid=none,tick label style={font=\scriptsize},label style={font=\small},title style={font=\small},legend style={font=\scriptsize,draw=none,fill=white},tick align=inside]
\nextgroupplot[title={(a) State preservation},xlabel={$\eta_{\rm L}=\eta_{\rm R}$},ylabel={$1-\mathcal F_{\rm pres}$},ymode=log,xmin=0,xmax=1,legend pos=south west]
\addplot[very thick,blue!75!black] coordinates {
(0,0.00274202819989)
(0.01,0.00273421024182)
(0.02,0.00272614071804)
(0.03,0.00271781215435)
(0.04,0.0027092168866)
(0.05,0.00270034705885)
(0.06,0.00269119462187)
(0.07,0.00268175133246)
(0.08,0.00267200875222)
(0.09,0.00266195824682)
(0.1,0.00265159098719)
(0.11,0.00264089794885)
(0.12,0.00262986991345)
(0.13,0.00261849747026)
(0.14,0.00260677101801)
(0.15,0.00259468076854)
(0.16,0.00258221674849)
(0.17,0.00256936880541)
(0.18,0.00255612661168)
(0.19,0.00254247967061)
(0.2,0.00252841732339)
(0.21,0.00251392875639)
(0.22,0.00249900301064)
(0.23,0.00248362899193)
(0.24,0.00246779548119)
(0.25,0.00245149114801)
(0.26,0.0024347045637)
(0.27,0.00241742421774)
(0.28,0.00239963853399)
(0.29,0.00238133588909)
(0.3,0.00236250463395)
(0.31,0.00234313311569)
(0.32,0.0023232097012)
(0.33,0.00230272280503)
(0.34,0.00228166091647)
(0.35,0.00226001263167)
(0.36,0.0022377666875)
(0.37,0.00221491199584)
(0.38,0.00219143768536)
(0.39,0.00216733314057)
(0.4,0.00214258804858)
(0.41,0.00211719244719)
(0.42,0.00209113677525)
(0.43,0.00206441192898)
(0.44,0.00203700932088)
(0.45,0.00200892094144)
(0.46,0.00198013942791)
(0.47,0.00195065813377)
(0.48,0.0019204712067)
(0.49,0.00188957366644)
(0.5,0.00185796149339)
(0.51,0.00182563171651)
(0.52,0.00179258251181)
(0.53,0.00175881330203)
(0.54,0.00172432486631)
(0.55,0.00168911945155)
(0.56,0.00165320089368)
(0.57,0.00161657474486)
(0.58,0.00157924840371)
(0.59,0.00154123125883)
(0.6,0.00150253483231)
(0.61,0.00146317293802)
(0.62,0.00142316183943)
(0.63,0.00138252042086)
(0.64,0.00134127036507)
(0.65,0.00129943633361)
(0.66,0.00125704616381)
(0.67,0.00121413106457)
(0.68,0.00117072582301)
(0.69,0.00112686901916)
(0.7,0.00108260324614)
(0.71,0.0010379753342)
(0.72,0.000993036587341)
(0.73,0.000947843015229)
(0.74,0.000902455577581)
(0.75,0.000856940427969)
(0.76,0.000811369158537)
(0.77,0.000765819050712)
(0.78,0.0007203733136)
(0.79,0.000675121332042)
(0.8,0.000630158897553)
(0.81,0.000585588430861)
(0.82,0.000541519189223)
(0.83,0.000498067441756)
(0.84,0.000455356611916)
(0.85,0.000413517354514)
(0.86,0.000372687557222)
(0.87,0.000333012212302)
(0.88,0.000294643134787)
(0.89,0.00025773845834)
(0.9,0.000222461882176)
(0.91,0.000188981679633)
(0.92,0.000157469585377)
(0.93,0.000128099935885)
(0.94,0.000101049792165)
(0.95,7.65013087221e-05)
(0.955,6.5225066582e-05)
(0.956,6.30529282351e-05)
(0.957,6.0909001092e-05)
(0.958,5.8793512185e-05)
(0.959,5.67066921791e-05)
(0.96,5.4648779076e-05)
(0.961,5.2620016842e-05)
(0.962,5.0620658594e-05)
(0.963,4.8650961145e-05)
(0.964,4.671119654e-05)
(0.965,4.4801644553e-05)
(0.966,4.2922597121e-05)
(0.967,4.1074361245e-05)
(0.968,3.9257259101e-05)
(0.969,3.747163168e-05)
(0.97,3.5717841115e-05)
(0.971,3.3996273982e-05)
(0.972,3.2307343915e-05)
(0.973,3.065149683e-05)
(0.974,2.9029214456e-05)
(0.975,2.7441020412e-05)
(0.976,2.5887485043e-05)
(0.977,2.4369230248e-05)
(0.978,2.2886940509e-05)
(0.979,2.14413631791e-05)
(0.98,2.0033322857e-05)
(0.981,1.8663722129e-05)
(0.982,1.7333555272e-05)
(0.983,1.60439149e-05)
(0.984,1.4796001513e-05)
(0.985,1.3591136404e-05)
(0.986,1.24307784281e-05)
(0.987,1.1316551395e-05)
(0.988,1.0250274773e-05)
(0.989,9.23403115305e-06)
(0.99,8.27023253402e-06)
(0.991,7.36173622595e-06)
(0.992,6.51196427204e-06)
(0.993,5.72502510199e-06)
(0.994,5.00571983297e-06)
(0.995,4.35921821496e-06)
(0.997,3.29458433301e-06)
(0.998,2.83859407801e-06)
(0.999,2.172052604e-06)
(0.9995,1.31928295499e-06)
};
\addlegendentry{biased}
\addplot[very thick,orange!85!black] coordinates {
(0,0.00262428672049)
(0.01,0.00261637677026)
(0.02,0.00260821121398)
(0.03,0.00259978240599)
(0.04,0.00259108250507)
(0.05,0.00258210347263)
(0.06,0.00257283707167)
(0.07,0.0025632748656)
(0.08,0.00255340821798)
(0.09,0.00254322829244)
(0.1,0.0025327260529)
(0.11,0.00252189226435)
(0.12,0.00251071749564)
(0.13,0.00249919212022)
(0.14,0.00248730632002)
(0.15,0.00247505008806)
(0.16,0.00246241323538)
(0.17,0.00244938539384)
(0.18,0.00243595602466)
(0.19,0.00242211442468)
(0.2,0.00240784973588)
(0.21,0.00239315095561)
(0.22,0.00237800694687)
(0.23,0.00236240645247)
(0.24,0.00234633810848)
(0.25,0.0023297904608)
(0.26,0.0023127519832)
(0.27,0.00229521109642)
(0.28,0.00227715619115)
(0.29,0.00225857565082)
(0.3,0.00223945787925)
(0.31,0.00221979132849)
(0.32,0.00219956453142)
(0.33,0.00217876613509)
(0.34,0.00215738493841)
(0.35,0.0021354099323)
(0.36,0.0021128303446)
(0.37,0.00208963568531)
(0.38,0.00206581579857)
(0.39,0.00204136091783)
(0.4,0.00201626172318)
(0.41,0.00199050940508)
(0.42,0.00196409573137)
(0.43,0.00193701311879)
(0.44,0.00190925471003)
(0.45,0.001880814454)
(0.46,0.00185168719267)
(0.47,0.00182186875168)
(0.48,0.00179135603794)
(0.49,0.00176014713955)
(0.5,0.0017282414339)
(0.51,0.00169563969944)
(0.52,0.00166234423237)
(0.53,0.00162835896945)
(0.54,0.00159368961503)
(0.55,0.00155834377253)
(0.56,0.00152233108051)
(0.57,0.00148566335302)
(0.58,0.00144835472216)
(0.59,0.00141042178462)
(0.6,0.00137188374916)
(0.61,0.00133276258566)
(0.62,0.00129308317461)
(0.63,0.00125287345434)
(0.64,0.00121216456648)
(0.65,0.00117099099878)
(0.66,0.00112939071919)
(0.67,0.0010874053059)
(0.68,0.00104508006643)
(0.69,0.00100246414637)
(0.7,0.00095961062361)
(0.71,0.000916576589329)
(0.72,0.000873423210087)
(0.73,0.000830215772044)
(0.74,0.000787023704681)
(0.75,0.000743920582845)
(0.76,0.000700984109592)
(0.77,0.000658296074713)
(0.78,0.000615942300955)
(0.79,0.000574012571914)
(0.8,0.000532600556339)
(0.81,0.000491803733334)
(0.82,0.000451723336117)
(0.83,0.000412464329146)
(0.84,0.000374135443064)
(0.85,0.000336849298865)
(0.86,0.000300722652049)
(0.87,0.000265876800187)
(0.88,0.000232438195968)
(0.89,0.000200539310804)
(0.9,0.000170319774764)
(0.91,0.000141927814701)
(0.92,0.000115521954869)
(0.93,9.1272900565e-05)
(0.94,6.9365447988e-05)
(0.95,5.0000214165e-05)
(0.955,4.1338301931e-05)
(0.956,3.9691213393e-05)
(0.957,3.8073104759e-05)
(0.958,3.6484215755e-05)
(0.959,3.4924787644e-05)
(0.96,3.3395063844e-05)
(0.961,3.18952898199e-05)
(0.962,3.0425713196e-05)
(0.963,2.8986584095e-05)
(0.964,2.7578154741e-05)
(0.965,2.6200680088e-05)
(0.966,2.48544176989e-05)
(0.967,2.3539627357e-05)
(0.968,2.2256572535e-05)
(0.969,2.100551949e-05)
(0.97,1.9786738059e-05)
(0.971,1.8600501772e-05)
(0.972,1.744708808e-05)
(0.973,1.6326779204e-05)
(0.974,1.5239862367e-05)
(0.975,1.4186630021e-05)
(0.976,1.31673809231e-05)
(0.977,1.21824213321e-05)
(0.978,1.1232063644e-05)
(0.979,1.0316630028e-05)
(0.98,9.43645163298e-06)
(0.981,8.59186995195e-06)
(0.982,7.78323875095e-06)
(0.983,7.01092486899e-06)
(0.984,6.27531026198e-06)
(0.985,5.57679541302e-06)
(0.986,4.91580109596e-06)
(0.987,4.292773776e-06)
(0.988,3.70819162998e-06)
(0.989,3.16257114896e-06)
(0.99,2.65647741604e-06)
(0.991,2.19053400696e-06)
(0.992,1.76543560304e-06)
(0.993,1.38195953203e-06)
(0.994,1.04098589104e-06)
(0.995,7.4355063695e-07)
(0.997,2.85560491031e-07)
(0.998,1.31016681948e-07)
(0.999,3.35427320364e-08)
(0.9995,8.4401110545e-09)
};
\addlegendentry{identical targets}
\nextgroupplot[title={(b) Thermal renewal},xlabel={$\eta_{\rm L}=\eta_{\rm R}$},ylabel={$\mathcal F_{\rm ref}$},xmin=0,xmax=1,ymin=0,ymax=1.04]
\addplot[very thick,blue!75!black] coordinates {
(0,1)
(0.01,0.999999721399)
(0.02,0.999998866184)
(0.03,0.999997404216)
(0.04,0.999995303939)
(0.05,0.999992532316)
(0.06,0.999989054764)
(0.07,0.999984835079)
(0.08,0.999979835363)
(0.09,0.99997401595)
(0.1,0.99996733532)
(0.11,0.999959750019)
(0.12,0.999951214571)
(0.13,0.999941681384)
(0.14,0.999931100655)
(0.15,0.999919420273)
(0.16,0.999906585715)
(0.17,0.999892539935)
(0.18,0.999877223256)
(0.19,0.999860573249)
(0.2,0.999842524614)
(0.21,0.999823009052)
(0.22,0.999801955136)
(0.23,0.999779288169)
(0.24,0.999754930047)
(0.25,0.999728799108)
(0.26,0.99970080998)
(0.27,0.999670873419)
(0.28,0.99963889615)
(0.29,0.999604780687)
(0.3,0.999568425163)
(0.31,0.999529723144)
(0.32,0.999488563434)
(0.33,0.999444829886)
(0.34,0.999398401191)
(0.35,0.999349150674)
(0.36,0.999296946072)
(0.37,0.999241649315)
(0.38,0.99918311629)
(0.39,0.99912119661)
(0.4,0.999055733365)
(0.41,0.998986562876)
(0.42,0.998913514436)
(0.43,0.998836410047)
(0.44,0.998755064155)
(0.45,0.998669283372)
(0.46,0.998578866202)
(0.47,0.998483602753)
(0.48,0.998383274455)
(0.49,0.998277653767)
(0.5,0.998166503889)
(0.51,0.998049578461)
(0.52,0.997926621279)
(0.53,0.997797365997)
(0.54,0.997661535839)
(0.55,0.997518843311)
(0.56,0.997368989922)
(0.57,0.997211665907)
(0.58,0.997046549963)
(0.59,0.99687330899)
(0.6,0.996691597842)
(0.61,0.996501059094)
(0.62,0.99630132281)
(0.63,0.99609200633)
(0.64,0.995872714057)
(0.65,0.995643037241)
(0.66,0.995402553765)
(0.67,0.995150827905)
(0.68,0.994887410046)
(0.69,0.994611836346)
(0.7,0.994323628288)
(0.71,0.994022292085)
(0.72,0.993707317875)
(0.73,0.993378178588)
(0.74,0.993034328401)
(0.75,0.992675200581)
(0.76,0.992300204508)
(0.77,0.99190872159)
(0.78,0.991500099639)
(0.79,0.991073645191)
(0.8,0.990628613001)
(0.81,0.990164191663)
(0.82,0.989679483916)
(0.83,0.989173479493)
(0.84,0.988645017461)
(0.85,0.988092733416)
(0.86,0.987514984425)
(0.87,0.986909740416)
(0.88,0.986274423452)
(0.89,0.985605663249)
(0.9,0.984898912661)
(0.91,0.984147818302)
(0.92,0.983343140042)
(0.93,0.982470785352)
(0.94,0.981507964353)
(0.95,0.980414917879)
(0.955,0.979798043723)
(0.956,0.979667318044)
(0.957,0.979533794861)
(0.958,0.979397297676)
(0.959,0.979257633052)
(0.96,0.979114588429)
(0.961,0.978967929587)
(0.962,0.978817397656)
(0.963,0.978662705609)
(0.964,0.978503534084)
(0.965,0.978339526408)
(0.966,0.978170282621)
(0.967,0.977995352242)
(0.968,0.977814225462)
(0.969,0.977626322327)
(0.97,0.977430979367)
(0.971,0.977227432904)
(0.972,0.977014798068)
(0.973,0.976792042149)
(0.974,0.976557950462)
(0.975,0.97631108219)
(0.976,0.976049712671)
(0.977,0.975771757157)
(0.978,0.975474668961)
(0.979,0.975155301704)
(0.98,0.974809720599)
(0.981,0.974432940244)
(0.982,0.97401855468)
(0.983,0.973558206577)
(0.984,0.97304081126)
(0.985,0.972451398483)
(0.986,0.97176934269)
(0.987,0.970965586171)
(0.988,0.96999814745)
(0.989,0.968804596177)
(0.99,0.967288918629)
(0.991,0.965297462099)
(0.992,0.962572287376)
(0.993,0.958654266469)
(0.994,0.952664019391)
(0.995,0.942751227484)
(0.997,0.885513915594)
(0.998,0.781913593002)
(0.999,0.431750235609)
(0.9995,0.109688739378)
};
\addplot[very thick,orange!85!black] coordinates {
(0,1)
(0.01,0.999999733413)
(0.02,0.999998915263)
(0.03,0.999997517018)
(0.04,0.999995508836)
(0.05,0.999992859514)
(0.06,0.999989536421)
(0.07,0.999985505444)
(0.08,0.999980730918)
(0.09,0.999975175566)
(0.1,0.999968800427)
(0.11,0.999961564784)
(0.12,0.999953426097)
(0.13,0.999944339921)
(0.14,0.999934259831)
(0.15,0.999923137342)
(0.16,0.999910921824)
(0.17,0.99989756042)
(0.18,0.999882997954)
(0.19,0.999867176846)
(0.2,0.999850037014)
(0.21,0.999831515784)
(0.22,0.999811547792)
(0.23,0.999790064884)
(0.24,0.999766996018)
(0.25,0.999742267159)
(0.26,0.999715801179)
(0.27,0.99968751775)
(0.28,0.999657333239)
(0.29,0.999625160602)
(0.3,0.999590909279)
(0.31,0.999554485087)
(0.32,0.999515790117)
(0.33,0.999474722628)
(0.34,0.999431176945)
(0.35,0.999385043367)
(0.36,0.999336208062)
(0.37,0.999284552984)
(0.38,0.999229955786)
(0.39,0.999172289739)
(0.4,0.999111423665)
(0.41,0.99904722187)
(0.42,0.998979544102)
(0.43,0.998908245503)
(0.44,0.998833176589)
(0.45,0.998754183241)
(0.46,0.998671106713)
(0.47,0.998583783665)
(0.48,0.998492046212)
(0.49,0.998395722004)
(0.5,0.998294634334)
(0.51,0.998188602273)
(0.52,0.998077440848)
(0.53,0.997960961248)
(0.54,0.997838971085)
(0.55,0.997711274687)
(0.56,0.997577673452)
(0.57,0.997437966254)
(0.58,0.997291949904)
(0.59,0.997139419677)
(0.6,0.996980169911)
(0.61,0.99681399467)
(0.62,0.996640688492)
(0.63,0.996460047212)
(0.64,0.99627186887)
(0.65,0.996075954704)
(0.66,0.995872110236)
(0.67,0.995660146437)
(0.68,0.995439880975)
(0.69,0.995211139549)
(0.7,0.994973757274)
(0.71,0.994727580129)
(0.72,0.99447246641)
(0.73,0.994208288182)
(0.74,0.993934932652)
(0.75,0.993652303405)
(0.76,0.993360321396)
(0.77,0.993058925567)
(0.78,0.992748072902)
(0.79,0.992427737675)
(0.8,0.992097909563)
(0.81,0.99175859019)
(0.82,0.991409787534)
(0.83,0.991051507437)
(0.84,0.990683741251)
(0.85,0.990306448374)
(0.86,0.989919532071)
(0.87,0.989522806563)
(0.88,0.989115952872)
(0.89,0.988698460269)
(0.9,0.988269549398)
(0.91,0.987828072016)
(0.92,0.987372380434)
(0.93,0.986900156147)
(0.94,0.986408179547)
(0.95,0.985892006928)
(0.955,0.985623012019)
(0.956,0.985568224988)
(0.957,0.985513089389)
(0.958,0.98545759646)
(0.959,0.985401737174)
(0.96,0.985345502233)
(0.961,0.985288882074)
(0.962,0.985231866875)
(0.963,0.985174446559)
(0.964,0.985116610816)
(0.965,0.985058349114)
(0.966,0.984999650732)
(0.967,0.984940504791)
(0.968,0.984880900298)
(0.969,0.984820826206)
(0.97,0.984760271478)
(0.971,0.984699225176)
(0.972,0.984637676559)
(0.973,0.984575615209)
(0.974,0.984513031175)
(0.975,0.984449915152)
(0.976,0.984386258681)
(0.977,0.984322054401)
(0.978,0.984257296331)
(0.979,0.984191980215)
(0.98,0.984126103917)
(0.981,0.984059667901)
(0.982,0.983992675754)
(0.983,0.983925134794)
(0.984,0.983857056678)
(0.985,0.983788457953)
(0.986,0.983719360369)
(0.987,0.983649790653)
(0.988,0.983579779345)
(0.989,0.983509358191)
(0.99,0.983438555961)
(0.991,0.983367393821)
(0.992,0.983295884861)
(0.993,0.983224049325)
(0.994,0.983151968075)
(0.995,0.983079909792)
(0.997,0.982939619154)
(0.998,0.982876397586)
(0.999,0.982824876026)
(0.9995,0.982805843657)
};
\nextgroupplot[title={(c) High-retention cut},xlabel={$\eta_{\rm R}$},ylabel={$\mathcal F$},xmin=0.97,xmax=1,ymode=log,xtick={0.97,0.98,0.99,1},legend pos=south west]
\addplot[very thick,teal!80!black] coordinates {
(0.97,0.998712880167)
(0.9705,0.998713828578)
(0.971,0.998714777067)
(0.9715,0.998715725511)
(0.972,0.998716673779)
(0.9725,0.998717621728)
(0.973,0.998718569208)
(0.9735,0.998719516054)
(0.974,0.998720462091)
(0.9745,0.998721407129)
(0.975,0.998722350968)
(0.9755,0.998723293391)
(0.976,0.998724234167)
(0.9765,0.998725173046)
(0.977,0.998726109764)
(0.9775,0.998727044036)
(0.978,0.998727975559)
(0.9785,0.998728904007)
(0.979,0.998729829033)
(0.9795,0.998730750264)
(0.98,0.998731667303)
(0.9805,0.998732579722)
(0.981,0.998733487066)
(0.9815,0.998734388846)
(0.982,0.998735284539)
(0.9825,0.998736173582)
(0.983,0.998737055375)
(0.9835,0.99873792927)
(0.984,0.998738794578)
(0.9845,0.998739650559)
(0.985,0.998740496421)
(0.9855,0.998741331326)
(0.986,0.998742154383)
(0.9865,0.998742964662)
(0.987,0.998743761197)
(0.9875,0.998744543012)
(0.988,0.998745309147)
(0.9885,0.99874605871)
(0.989,0.998746790955)
(0.9895,0.998747505393)
(0.99,0.998748201978)
(0.9905,0.998748881372)
(0.991,0.998749545363)
(0.9915,0.998750197498)
(0.992,0.998750844062)
(0.9925,0.998751495584)
(0.993,0.998752169171)
(0.9935,0.998752892128)
(0.994,0.998753707595)
(0.9945,0.998754683416)
(0.995,0.998755926323)
(0.9951,0.998756220351)
(0.9952,0.99875653324)
(0.9953,0.998756866899)
(0.9954,0.998757223424)
(0.9955,0.998757605124)
(0.9956,0.998758014538)
(0.9957,0.998758454463)
(0.9958,0.998758927983)
(0.9959,0.998759438497)
(0.996,0.998759989762)
(0.9961,0.998760585928)
(0.9962,0.998761231583)
(0.9963,0.998761931809)
(0.9964,0.998762692238)
(0.9965,0.998763519116)
(0.9966,0.998764419379)
(0.9967,0.998765400738)
(0.9968,0.998766471772)
(0.9969,0.998767642039)
(0.997,0.998768922199)
(0.9971,0.998770324153)
(0.9972,0.998771861204)
(0.9973,0.998773548243)
(0.9974,0.998775401957)
(0.9975,0.998777441071)
(0.9976,0.998779686628)
(0.9977,0.99878216231)
(0.9978,0.998784894817)
(0.9979,0.9987879143)
(0.998,0.99879125488)
(0.9981,0.99879495526)
(0.9982,0.998799059453)
(0.9983,0.998803617664)
(0.9984,0.998808687373)
(0.9985,0.998814334673)
(0.9986,0.998820635963)
(0.9987,0.998827680128)
(0.9988,0.998835571419)
(0.9989,0.99884443334)
(0.999,0.998854414073)
(0.9991,0.998865694313)
(0.9992,0.998878499024)
(0.9993,0.998893115908)
(0.9994,0.998909926059)
(0.9995,0.998929458289)
(0.9996,0.998952493842)
(0.9997,0.998980292662)
(0.9998,0.999015175341)
(0.9999,0.999062605476)
(1,0.999157149129)
};
\addlegendentry{$\mathcal F_{\rm pres}$}
\addplot[very thick,violet!85!black] coordinates {
(0.97,0.961591795521)
(0.9705,0.960996847412)
(0.971,0.960382500739)
(0.9715,0.959747765505)
(0.972,0.959091582431)
(0.9725,0.958412816756)
(0.973,0.957710251354)
(0.9735,0.956982579074)
(0.974,0.956228394212)
(0.9745,0.955446182985)
(0.975,0.954634312873)
(0.9755,0.953791020668)
(0.976,0.952914399033)
(0.9765,0.95200238137)
(0.977,0.951052724708)
(0.9775,0.950062990341)
(0.978,0.949030521826)
(0.9785,0.947952419929)
(0.979,0.946825514016)
(0.9795,0.945646329274)
(0.98,0.944411049043)
(0.9805,0.943115471407)
(0.981,0.941754958971)
(0.9815,0.940324380594)
(0.982,0.938818043502)
(0.9825,0.937229613937)
(0.983,0.935552023983)
(0.9835,0.933777361735)
(0.984,0.931896741208)
(0.9845,0.92990014752)
(0.985,0.927776251719)
(0.9855,0.925512188123)
(0.986,0.92309328513)
(0.9865,0.920502737956)
(0.987,0.917721208446)
(0.9875,0.914726332739)
(0.988,0.911492111595)
(0.9885,0.907988150133)
(0.989,0.904178702488)
(0.9895,0.900021461346)
(0.99,0.895466010474)
(0.9905,0.890451827236)
(0.991,0.884905677458)
(0.9915,0.878738179636)
(0.992,0.871839218603)
(0.9925,0.864071742617)
(0.993,0.855263253531)
(0.9935,0.845193949212)
(0.994,0.833579918553)
(0.9945,0.820048879412)
(0.995,0.804104431537)
(0.9951,0.800570526339)
(0.9952,0.796907478486)
(0.9953,0.793108401422)
(0.9954,0.789165932593)
(0.9955,0.785072193232)
(0.9956,0.780818744128)
(0.9957,0.776396536924)
(0.9958,0.771795860407)
(0.9959,0.767006281191)
(0.996,0.762016578121)
(0.9961,0.756814669601)
(0.9962,0.751387532984)
(0.9963,0.745721114962)
(0.9964,0.739800231814)
(0.9965,0.733608458127)
(0.9966,0.727128002426)
(0.9967,0.720339567877)
(0.9968,0.713222195924)
(0.9969,0.705753090365)
(0.997,0.697907418922)
(0.9971,0.68965808884)
(0.9972,0.680975492416)
(0.9973,0.671827217582)
(0.9974,0.662177717719)
(0.9975,0.651987933751)
(0.9976,0.641214860162)
(0.9977,0.629811044853)
(0.9978,0.617724010714)
(0.9979,0.604895584233)
(0.998,0.591261113417)
(0.9981,0.576748553748)
(0.9982,0.561277396668)
(0.9983,0.544757410594)
(0.9984,0.527087159868)
(0.9985,0.50815226366)
(0.9986,0.487823356826)
(0.9987,0.465953723143)
(0.9988,0.442376598699)
(0.9989,0.416902211589)
(0.999,0.389314781158)
(0.9991,0.359370050802)
(0.9992,0.326794713141)
(0.9993,0.291290879808)
(0.9994,0.252553006009)
(0.9995,0.210315356164)
(0.9996,0.16447689869)
(0.9997,0.115435999669)
(0.9998,0.065053836893)
(0.9999,0.0197671366814)
(1,1.75812396365e-08)
};
\addlegendentry{$\mathcal F_{\rm ref}$}
\end{groupplot}
\end{tikzpicture}
\caption{Full Gaussian many-body fidelity for the parameters of Eq.~\eqref{eq:params}, with $N_{\rm L}=N_{\rm R}=28$. (a) Infidelity across the refresh stroke and (b) fidelity with the thermal product reference, for symmetric attenuation. Blue: original chemical bias; orange: identical targets ($\mu_{\rm L}=\mu_{\rm R}=0$), at the same temperature. The diagonal stops at $\eta=0.9995$, excluding the nonattractive unitary corner. (c) The biased high-retention cut at $\eta_{\rm L}=0.42529$ includes the singly unrefreshed endpoint, whose fixed point is attractive for this finite benchmark. The logarithmic scales in (a) and (c) resolve distinct fidelity ranges. All data are calculated directly from Eq.~\eqref{eq:fidexact}.}\label{fig:fidelity}
\end{figure*}

\subsection{Coherence and energetic operating landscapes}
\begin{figure*}[t]
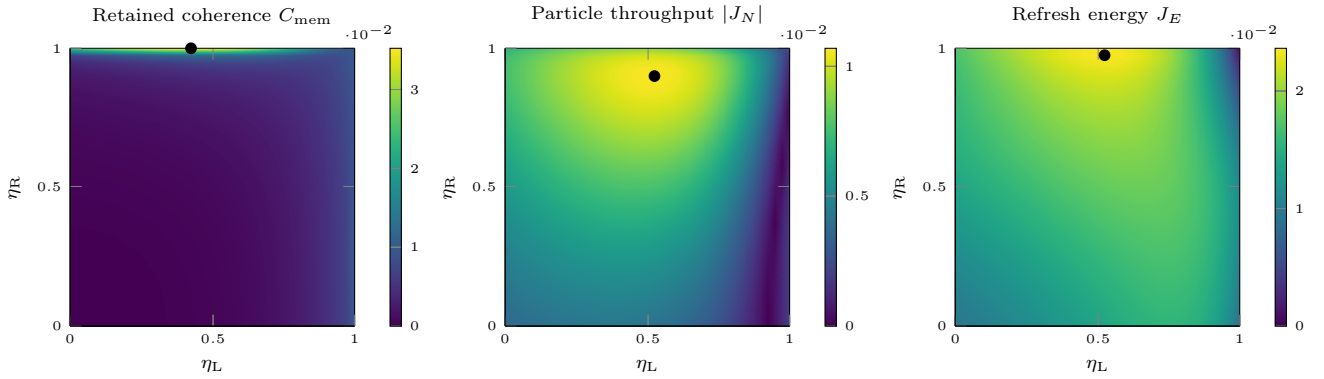

\centering

\caption{Stationary operating landscapes for the Gaussian partial-refresh protocol with $N_{\rm L}=N_{\rm R}=28$ and the parameters of Eq.~\eqref{eq:params}. From left to right: retained system--bath coherence $C_{\rm mem}$, particle-current magnitude $|J_N|$, and signed energy-transfer rate $J_E$ into the outgoing thermal ancillas. The black marker in each panel denotes the maximum on the displayed $41\times41$ parameter grid. All stationary covariances satisfy $0\le C\le\Id$. The parameter range is restricted to $0\le\eta_a\le0.995$; the nonrefreshing endpoint $\eta_a=1$ is not included.}\label{fig:maps}
\end{figure*}

The operating diagrams below are obtained from the microscopically defined Gaussian partial-refresh channel of Eq.~\eqref{eq:refresh}. Figure~\ref{fig:maps} displays the scanned $41\times41$ operating grid in $0\le\eta_a\le0.995$. Grid maxima, with their scan resolution understood, occur at
\begin{align}
\mc\text{ maximum}:&(\eta_{\mathrm L},\eta_{\mathrm R})=(0.4229,0.9950),\nonumber\\
|\jn|\text{ maximum}:&(0.5224,0.8955),\nonumber\\
\je\text{ maximum}:&(0.5224,0.9701).\label{eq:optloc}
\end{align}
The corresponding optimum values are $\mc=0.03523$, $|\jn|=0.01069$, and $\je=0.02363$ in the stated units. For this benchmark, the coherence-optimal and signed-energy-optimal regions differ; the particle-current maximum is retained as a separate comparison point. These maximizers are specific to the finite retention grid and do not establish a parameter-independent optimization theorem. Continuous \emph{local} refinement constrained to the same scan square yields, respectively,
\begin{align}
(\eta_{\rm L},\eta_{\rm R})_{\rm mem}^{\rm loc}&=(0.42529,0.99500),\nonumber\\
(\eta_{\rm L},\eta_{\rm R})_{\rm curr}^{\rm loc}&=(0.51759,0.89861),\nonumber\\
(\eta_{\rm L},\eta_{\rm R})_{\rm energy}^{\rm loc}&=(0.52424,0.98317),\label{eq:refined}
\end{align}
with values $C_{\rm mem}=0.0352265$, $|J_N|=0.0106939$, and $J_E=0.0237189$. The three peak coordinates reflect the chosen bias, near-band-edge level, finite cycle time, and particular Hilbert--Schmidt definition of memory. They should not be interpreted as an invariant three-way thermodynamic inequality. In fact, for a different reference bias the signed energy observable may change sign, so the location of its largest positive value need not track the largest magnitude or any minimum-cost operating point.

The markers in Fig.~\ref{fig:maps} refer to the grid maxima in Eq.~\eqref{eq:optloc}, whereas Eq.~\eqref{eq:refined} gives the constrained local refinements. The memory optimum within the specified scan region occurs at its upper $\eta_{\rm R}=0.995$ boundary.

\begin{figure*}[t]
\centering
\begin{tikzpicture}
\begin{axis}[width=0.43\textwidth,height=0.265\textwidth,xmin=0,xmax=1,ymin=0,ymax=1.08,
xlabel={$\eta_{\rm L}=\eta_{\rm R}=\eta$},ylabel={normalized observable},
label style={font=\small},tick label style={font=\scriptsize},legend style={font=\scriptsize,draw=none,fill=none,at={(0.03,0.97)},anchor=north west},
no markers,thick]
\addplot[very thick,blue] table[row sep=crcr] {
0 3.20365918583e-33\\
0.01 1.41267614598e-05\\
0.02 5.74909144078e-05\\
0.03 0.000131619991185\\
0.04 0.00023811300466\\
0.05 0.000378643694637\\
0.06 0.000554963912166\\
0.07 0.000768907146459\\
0.08 0.00102239219919\\
0.09 0.00131742701096\\
0.1 0.00165611264464\\
0.11 0.00204064743044\\
0.12 0.00247333127735\\
0.13 0.00295657015554\\
0.14 0.00349288075428\\
0.15 0.00408489531976\\
0.16 0.00473536667682\\
0.17 0.00544717343874\\
0.18 0.00622332540854\\
0.19 0.00706696917501\\
0.2 0.00798139390654\\
0.21 0.00897003734475\\
0.22 0.0100364919999\\
0.23 0.0111845115489\\
0.24 0.0124180174363\\
0.25 0.0137411056772\\
0.26 0.015158053861\\
0.27 0.0166733283513\\
0.28 0.0182915916802\\
0.29 0.0200177101275\\
0.3 0.0218567614798\\
0.31 0.0238140429579\\
0.32 0.0258950792996\\
0.33 0.0281056309852\\
0.34 0.0304517025838\\
0.35 0.0329395512029\\
0.36 0.035575695012\\
0.37 0.0383669218138\\
0.38 0.0413202976252\\
0.39 0.0444431752302\\
0.4 0.0477432026577\\
0.41 0.0512283315305\\
0.42 0.0549068252249\\
0.43 0.0587872667716\\
0.44 0.0628785664198\\
0.45 0.0671899687722\\
0.46 0.0717310593935\\
0.47 0.0765117707738\\
0.48 0.0815423875219\\
0.49 0.086833550641\\
0.5 0.0923962607257\\
0.51 0.0982418799\\
0.52 0.104382132292\\
0.53 0.11082910282\\
0.54 0.117595234049\\
0.55 0.124693320822\\
0.56 0.132136502389\\
0.57 0.139938251677\\
0.58 0.148112361343\\
0.59 0.156672926225\\
0.6 0.165634321742\\
0.61 0.175011177795\\
0.62 0.184818347673\\
0.63 0.195070871448\\
0.64 0.205783933306\\
0.65 0.216972812284\\
0.66 0.228652825821\\
0.67 0.24083926561\\
0.68 0.253547325216\\
0.69 0.266792019033\\
0.7 0.280588092221\\
0.71 0.294949921464\\
0.72 0.309891406563\\
0.73 0.325425853258\\
0.74 0.341565848078\\
0.75 0.358323126631\\
0.76 0.375708437613\\
0.77 0.393731405897\\
0.78 0.412400399641\\
0.79 0.43172240844\\
0.8 0.451702942443\\
0.81 0.472345966328\\
0.82 0.493653887625\\
0.83 0.515627626679\\
0.84 0.538266806708\\
0.85 0.561570118433\\
0.86 0.585535937025\\
0.87 0.610163303477\\
0.88 0.63545343373\\
0.89 0.661411996848\\
0.9 0.688052524793\\
0.91 0.715401510962\\
0.92 0.743506081049\\
0.93 0.772445701768\\
0.94 0.802350517211\\
0.95 0.833431282887\\
0.96 0.866031175609\\
0.97 0.900719442229\\
0.971 0.904335355708\\
0.972 0.907982364396\\
0.973 0.911661720418\\
0.974 0.915374722632\\
0.975 0.919122711295\\
0.976 0.922907059679\\
0.977 0.926729161423\\
0.978 0.930590411995\\
0.979 0.934492181975\\
0.98 0.938435779011\\
0.981 0.942422394052\\
0.982 0.946453025794\\
0.983 0.950528374715\\
0.984 0.95464869428\\
0.985 0.958813580421\\
0.986 0.963021668633\\
0.987 0.967270184561\\
0.988 0.971554244298\\
0.989 0.975865691009\\
0.99 0.980191006228\\
0.991 0.984507259767\\
0.992 0.988773701338\\
0.993 0.992913258224\\
0.994 0.996769526609\\
0.995 1\\
};
\addlegendentry{$C_{\mathrm{mem}}$}
\addplot[very thick,orange] table[row sep=crcr] {
0 0.435598995853\\
0.01 0.443300347733\\
0.02 0.451098755023\\
0.03 0.458994794457\\
0.04 0.466988974193\\
0.05 0.475081727151\\
0.06 0.483273403915\\
0.07 0.491564265119\\
0.08 0.499954473327\\
0.09 0.508444084359\\
0.1 0.517033038029\\
0.11 0.525721148281\\
0.12 0.534508092652\\
0.13 0.543393401071\\
0.14 0.552376443919\\
0.15 0.561456419323\\
0.16 0.570632339668\\
0.17 0.579903017228\\
0.18 0.589267048946\\
0.19 0.598722800248\\
0.2 0.608268387904\\
0.21 0.61790166185\\
0.22 0.627620185938\\
0.23 0.637421217579\\
0.24 0.647301686203\\
0.25 0.657258170513\\
0.26 0.667286874466\\
0.27 0.677383601953\\
0.28 0.687543730099\\
0.29 0.697762181177\\
0.3 0.708033393058\\
0.31 0.71835128817\\
0.32 0.728709240932\\
0.33 0.739100043601\\
0.34 0.749515870538\\
0.35 0.759948240833\\
0.36 0.7703879793\\
0.37 0.780825175786\\
0.38 0.791249142845\\
0.39 0.801648371742\\
0.4 0.812010486813\\
0.41 0.822322198226\\
0.42 0.832569253179\\
0.43 0.842736385596\\
0.44 0.85280726442\\
0.45 0.862764440584\\
0.46 0.872589292827\\
0.47 0.882261972479\\
0.48 0.891761347439\\
0.49 0.901064945561\\
0.5 0.910148897727\\
0.51 0.918987880908\\
0.52 0.927555061613\\
0.53 0.935822040131\\
0.54 0.943758796056\\
0.55 0.951333635689\\
0.56 0.95851314193\\
0.57 0.965262127417\\
0.58 0.971543591749\\
0.59 0.977318683746\\
0.6 0.982546669806\\
0.61 0.987184909604\\
0.62 0.991188840467\\
0.63 0.994511971987\\
0.64 0.997105892599\\
0.65 0.998920290023\\
0.66 0.999902987757\\
0.67 1\\
0.68 0.999155607646\\
0.69 0.997312458317\\
0.7 0.994411693616\\
0.71 0.990393107175\\
0.72 0.985195337316\\
0.73 0.97875609848\\
0.74 0.971012455898\\
0.75 0.961901148198\\
0.76 0.951358962912\\
0.77 0.93932316997\\
0.78 0.925732018292\\
0.79 0.910525300499\\
0.8 0.893644990379\\
0.81 0.875035957162\\
0.82 0.854646759581\\
0.83 0.832430521174\\
0.84 0.808345886033\\
0.85 0.782358051047\\
0.86 0.75443986625\\
0.87 0.724572988798\\
0.88 0.692749067461\\
0.89 0.658970922313\\
0.9 0.623253666149\\
0.91 0.585625686065\\
0.92 0.546129357177\\
0.93 0.504821280114\\
0.94 0.461771717926\\
0.95 0.417063168299\\
0.96 0.370794623365\\
0.97 0.323184800935\\
0.971 0.318378086851\\
0.972 0.313571189795\\
0.973 0.308768117439\\
0.974 0.30397400379\\
0.975 0.299195435542\\
0.976 0.294440875327\\
0.977 0.289721212159\\
0.978 0.2850504799\\
0.979 0.280446799195\\
0.98 0.275933619242\\
0.981 0.271541366035\\
0.982 0.26730964856\\
0.983 0.263290242133\\
0.984 0.25955117291\\
0.985 0.256182393885\\
0.986 0.253303813593\\
0.987 0.251076893336\\
0.988 0.249721815995\\
0.989 0.24954364282\\
0.99 0.250973515934\\
0.991 0.254636135703\\
0.992 0.261465427088\\
0.993 0.272913824513\\
0.994 0.291356342356\\
0.995 0.320934700744\\
};
\addlegendentry{$|J_N|$}
\addplot[very thick,teal!85!black] table[row sep=crcr] {
0 0.475919495013\\
0.01 0.48389647785\\
0.02 0.491952348877\\
0.03 0.500086646778\\
0.04 0.508298811683\\
0.05 0.51658817933\\
0.06 0.524953974973\\
0.07 0.533395307051\\
0.08 0.541911160604\\
0.09 0.550500390428\\
0.1 0.559161713981\\
0.11 0.567893704013\\
0.12 0.576694780929\\
0.13 0.585563204884\\
0.14 0.5944970676\\
0.15 0.603494283892\\
0.16 0.612552582919\\
0.17 0.621669499143\\
0.18 0.630842362997\\
0.19 0.640068291243\\
0.2 0.649344177052\\
0.21 0.658666679745\\
0.22 0.668032214254\\
0.23 0.677436940228\\
0.24 0.686876750844\\
0.25 0.696347261255\\
0.26 0.705843796697\\
0.27 0.715361380243\\
0.28 0.724894720161\\
0.29 0.734438196882\\
0.3 0.743985849557\\
0.31 0.753531362145\\
0.32 0.763068049041\\
0.33 0.772588840177\\
0.34 0.782086265558\\
0.35 0.79155243919\\
0.36 0.800979042333\\
0.37 0.810357306008\\
0.38 0.81967799267\\
0.39 0.828931376974\\
0.4 0.838107225502\\
0.41 0.847194775344\\
0.42 0.856182711387\\
0.43 0.86505914215\\
0.44 0.873811573978\\
0.45 0.882426883416\\
0.46 0.8908912875\\
0.47 0.899190311721\\
0.48 0.907308755378\\
0.49 0.915230653983\\
0.5 0.922939238366\\
0.51 0.930416890086\\
0.52 0.937645092685\\
0.53 0.944604378319\\
0.54 0.951274269221\\
0.55 0.957633213389\\
0.56 0.963658513888\\
0.57 0.96932625103\\
0.58 0.974611196671\\
0.59 0.979486719808\\
0.6 0.983924682557\\
0.61 0.98789532555\\
0.62 0.9913671417\\
0.63 0.994306737212\\
0.64 0.996678678645\\
0.65 0.998445324733\\
0.66 0.999566641613\\
0.67 1\\
0.68 0.999699952774\\
0.69 0.998617991352\\
0.7 0.996702279092\\
0.71 0.993897359919\\
0.72 0.990143840151\\
0.73 0.985378041442\\
0.74 0.979531622491\\
0.75 0.97253116694\\
0.76 0.964297734567\\
0.77 0.95474637237\\
0.78 0.943785581536\\
0.79 0.931316735342\\
0.8 0.917233441754\\
0.81 0.901420842624\\
0.82 0.883754838756\\
0.83 0.864101226244\\
0.84 0.842314723962\\
0.85 0.818237864006\\
0.86 0.791699704922\\
0.87 0.762514309752\\
0.88 0.730478903957\\
0.89 0.695371586318\\
0.9 0.656948399455\\
0.91 0.61493945736\\
0.92 0.569043640955\\
0.93 0.518921038693\\
0.94 0.464181672632\\
0.95 0.404367732254\\
0.96 0.338923419722\\
0.97 0.267137797524\\
0.971 0.259575888206\\
0.972 0.251939901559\\
0.973 0.244228584744\\
0.974 0.236440624845\\
0.975 0.228574643604\\
0.976 0.220629191431\\
0.977 0.212602740593\\
0.978 0.204493677388\\
0.979 0.196300293192\\
0.98 0.188020774261\\
0.981 0.179653190306\\
0.982 0.171195481998\\
0.983 0.162645447929\\
0.984 0.154000731938\\
0.985 0.145258812194\\
0.986 0.136416993904\\
0.987 0.127472407186\\
0.988 0.118422010108\\
0.989 0.109262591842\\
0.99 0.0999907602741\\
0.991 0.0906028787011\\
0.992 0.0810948869226\\
0.993 0.0714619135984\\
0.994 0.0616976051884\\
0.995 0.0517932928853\\
};
\addlegendentry{$J_E$}
\end{axis}\end{tikzpicture}\hspace{0.02\textwidth}
\begin{tikzpicture}
\begin{axis}[width=0.43\textwidth,height=0.265\textwidth,xmin=-2.1,xmax=2.1,
xlabel={bath eigenenergy $\omega$},ylabel={$|C_{0,ak}|^2$},
label style={font=\small},tick label style={font=\scriptsize},
legend style={font=\scriptsize,draw=none,fill=none,at={(0.04,0.96)},anchor=north west}]
\addplot[thick,blue,mark=*,mark size=.9pt] table[row sep=crcr] {
-1.988276 5.37006076017e-07\\
-1.953241 2.14940918103e-06\\
-1.895306 4.84069014763e-06\\
-1.815151 8.61431928535e-06\\
-1.713714 1.3469954427e-05\\
-1.592186 1.9397474447e-05\\
-1.451991 2.63680953939e-05\\
-1.294773 3.43213131408e-05\\
-1.122374 4.31452736655e-05\\
-0.9368169 5.26452075348e-05\\
-0.7402763 6.24867450257e-05\\
-0.5350567 7.20816879592e-05\\
-0.323564 8.03446446809e-05\\
-0.1082778 8.52096295399e-05\\
0.1082778 8.29663473883e-05\\
0.323564 6.86656128216e-05\\
0.5350567 4.13965901597e-05\\
0.7402763 1.32140309222e-05\\
0.9368169 5.05390182928e-07\\
1.122374 2.07141327684e-06\\
1.294773 6.72534697863e-06\\
1.451991 9.02281620702e-06\\
1.592186 8.80122568512e-06\\
1.713714 7.17238491504e-06\\
1.815151 5.0548604605e-06\\
1.895306 3.01845329203e-06\\
1.953241 1.39117378318e-06\\
1.988276 3.54695987877e-07\\
};
\addlegendentry{$a={\rm L}$}
\addplot[thick,orange,mark=square*,mark size=.9pt] table[row sep=crcr] {
-1.988276 1.34734532595e-06\\
-1.953241 5.40303164791e-06\\
-1.895306 1.21937953498e-05\\
-1.815151 2.17031507465e-05\\
-1.713714 3.3720610264e-05\\
-1.592186 4.74686918904e-05\\
-1.451991 6.07661844545e-05\\
-1.294773 6.8580763744e-05\\
-1.122374 6.24317848626e-05\\
-0.9368169 3.70778278004e-05\\
-0.7402763 7.31644957032e-06\\
-0.5350567 1.92978154211e-06\\
-0.323564 2.48429214357e-05\\
-0.1082778 5.84139328295e-05\\
0.1082778 9.19809574066e-05\\
0.323564 0.000124889139288\\
0.5350567 0.000159739382674\\
0.7402763 0.000199104821584\\
0.9368169 0.0002445962438\\
1.122374 0.000295987206516\\
1.294773 0.000349205437762\\
1.451991 0.000392962358414\\
1.592186 0.000406926144088\\
1.713714 0.000370146930623\\
1.815151 0.000282573738899\\
1.895306 0.000173778800121\\
1.953241 7.96496029245e-05\\
1.988276 1.99963829623e-05\\
};
\addlegendentry{$a={\rm R}$}
\end{axis}\end{tikzpicture}
\caption{Stationary observables for the Gaussian partial-refresh protocol. Left: retained coherence, particle-current magnitude, and signed energy-transfer rate along $\eta_{\rm L}=\eta_{\rm R}$, each divided by its maximum along that cut to compare the locations of the extrema. Right: mode-resolved coherence weights $|C_{0,ak}|^2$ for the left and right reservoirs at the two-parameter current optimum $(\eta_{\rm L},\eta_{\rm R})=(0.5224,0.8955)$.}\label{fig:cuts}
\end{figure*}
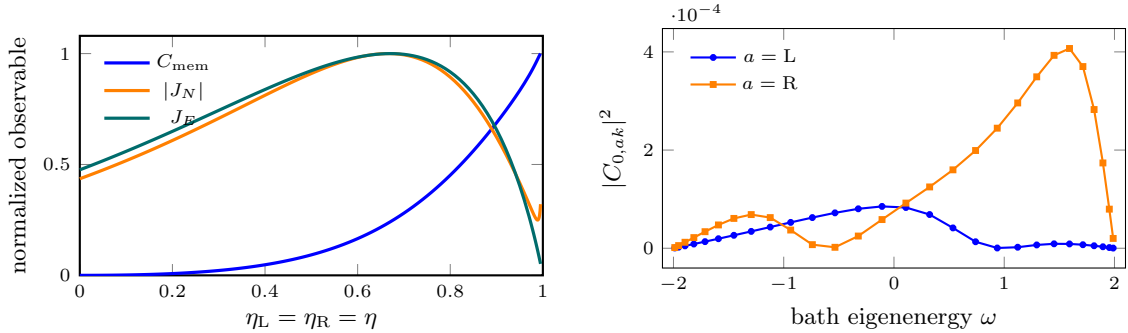

The symmetric-cut curves in Fig.~\ref{fig:cuts} show that the stored coherence and signed energy exchange respond differently to the physical reset. On the diagonal $\eta_{\rm L}=\eta_{\rm R}$, memory grows through most of the displayed range, while the energetic and particle-exchange diagnostics have intermediate maxima. This behavior is nonperturbative in the retention parameter and cannot be deduced by substituting $\eta_a=1$ into Eq.~\eqref{eq:gamma}. The side-resolved coherence spectra illustrate the nonuniform distribution of stored correlations over the bath modes at the marked current-selected comparison point.

\begin{figure*}[t]\centering
\begin{tikzpicture}\begin{axis}[width=0.31\textwidth,height=0.225\textwidth,xmin=12,xmax=72,
xlabel={chain length $N_{\rm L}=N_{\rm R}$},ylabel={$C_{\rm mem}$},
label style={font=\small},tick label style={font=\scriptsize},scaled y ticks=false,
legend style={at={(0.52,0.48)},anchor=center,font=\tiny,draw=none,fill=white},]
\addplot[thick,blue,mark=*,mark size=1.35pt] table[row sep=crcr] {
14 0.0328635029707\\
18 0.0350849526778\\
22 0.0351426454588\\
28 0.035225916824\\
34 0.0354501658799\\
42 0.0355108092787\\
56 0.0355594974844\\
70 0.0355715338201\\
};
\addlegendentry{memory point}
\addplot[thick,orange,mark=square*,mark size=1.35pt] table[row sep=crcr] {
14 0.00439075341365\\
18 0.00439070414653\\
22 0.00439070520984\\
28 0.00439070521109\\
34 0.00439070521414\\
42 0.00439070521413\\
56 0.00439070521413\\
70 0.00439070521413\\
};
\addlegendentry{current point}
\addplot[thick,teal!80!black,mark=triangle*,mark size=1.35pt] table[row sep=crcr] {
14 0.0116858526283\\
18 0.0116870713139\\
22 0.0116872884974\\
28 0.0116873461875\\
34 0.0116873558889\\
42 0.0116873576138\\
56 0.011687357784\\
70 0.0116873577868\\
};
\addlegendentry{energy point}
\end{axis}\end{tikzpicture}
\begin{tikzpicture}\begin{axis}[width=0.31\textwidth,height=0.225\textwidth,xmin=12,xmax=72,
xlabel={chain length $N_{\rm L}=N_{\rm R}$},ylabel={$|J_N|$},
label style={font=\small},tick label style={font=\scriptsize},scaled y ticks=false,
]
\addplot[thick,blue,mark=*,mark size=1.35pt] table[row sep=crcr] {
14 0.00904658065435\\
18 0.00873093112211\\
22 0.00879929185727\\
28 0.00893433027816\\
34 0.00892998770091\\
42 0.00894216370508\\
56 0.0089451649982\\
70 0.00894586551011\\
};
\addplot[thick,orange,mark=square*,mark size=1.35pt] table[row sep=crcr] {
14 0.0106928553712\\
18 0.0106927542347\\
22 0.0106927586346\\
28 0.0106927584629\\
34 0.010692758463\\
42 0.010692758463\\
56 0.010692758463\\
70 0.010692758463\\
};
\addplot[thick,teal!80!black,mark=triangle*,mark size=1.35pt] table[row sep=crcr] {
14 0.0100574498379\\
18 0.0100690818443\\
22 0.0100673320895\\
28 0.0100675084663\\
34 0.0100674961674\\
42 0.0100674967815\\
56 0.0100674967881\\
70 0.0100674967879\\
};
\end{axis}\end{tikzpicture}
\begin{tikzpicture}\begin{axis}[width=0.31\textwidth,height=0.225\textwidth,xmin=12,xmax=72,
xlabel={chain length $N_{\rm L}=N_{\rm R}$},ylabel={$J_E$},
label style={font=\small},tick label style={font=\scriptsize},scaled y ticks=false,
]
\addplot[thick,blue,mark=*,mark size=1.35pt] table[row sep=crcr] {
14 0.023477402283\\
18 0.0228876534888\\
22 0.0229941794902\\
28 0.0232379061893\\
34 0.0232277329999\\
42 0.0232480435797\\
56 0.0232524417153\\
70 0.0232533572756\\
};
\addplot[thick,orange,mark=square*,mark size=1.35pt] table[row sep=crcr] {
14 0.0224229192615\\
18 0.0224229776869\\
22 0.0224229758016\\
28 0.022422975861\\
34 0.0224229758624\\
42 0.0224229758624\\
56 0.0224229758624\\
70 0.0224229758624\\
};
\addplot[thick,teal!80!black,mark=triangle*,mark size=1.35pt] table[row sep=crcr] {
14 0.0236127217193\\
18 0.0236376853964\\
22 0.0236342459495\\
28 0.0236345703769\\
34 0.0236345492661\\
42 0.0236345515174\\
56 0.0236345516877\\
70 0.0236345516926\\
};
\end{axis}\end{tikzpicture}
\caption{Bath-size dependence of the stationary observables for the Gaussian partial-refresh protocol. The retention parameters are held fixed at the respective $N_a=28$ grid maxima listed in Eq.~\eqref{eq:optloc}. From left to right: retained coherence $C_{\rm mem}$, particle-current magnitude $|J_N|$, and signed energy-transfer rate $J_E$. The legend in the left panel applies to all three panels.}\label{fig:conv}\end{figure*}
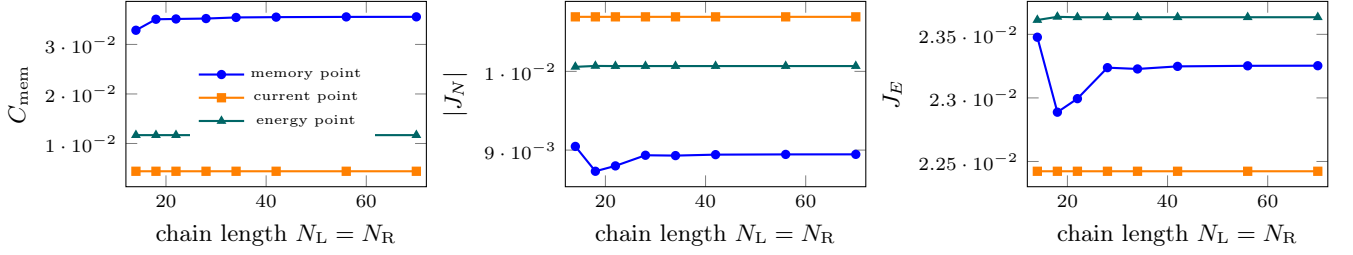

Figure~\ref{fig:conv} displays the bath-size dependence at fixed parameter pairs selected from the $N=28$ scan; these parameters are not reoptimized as the bath size changes. At the grid memory point $(0.422875,0.995)$ the finite-$N$ memory changes from $0.0352259$ at $N=28$ to approximately $0.03557$ at $N=70$ (about a one-percent upward shift); this modest change is nevertheless relevant when quoting high-precision limiting values. The particle-exchange diagnostic is more size-stable over this range. To assess the stability of the optimum locations, we also locally optimize the three observables separately for $N=14,28,48,64$ within $[0,0.995]^2$. The resulting local maximizers are tabulated in Table~\ref{tab:sizeopt}. The memory-optimal $\eta_{\rm L}$ shifts from $0.43534$ at $N=14$ to $0.42460$ at $N=64$ while remaining pinned to the imposed high-retention boundary. The energy maximum moves from that same boundary at $N=14$ to an interior point for the larger chains, while the particle-exchange optimum changes by less than $2\times10^{-5}$ in either parameter. Thus the numerical operating-point separation survives each reoptimization, but the smaller-size energy optimum is less stable. Neither a thermodynamic-limit theorem nor global optimization over the full retention square is claimed. Near the undamped boundary, the limits $N\to\infty$ and $\eta_a\to1$ require further analysis.
\begin{table}[t]
\caption{Constrained local maxima of retained coherence $C_{\rm mem}$, particle-current magnitude $|J_N|$, and signed energy-transfer rate $J_E$ for equal reservoir lengths $N_{\rm L}=N_{\rm R}=N_a$. Entries give the optimizing pairs $(\eta_{\rm L},\eta_{\rm R})$ within $0\le\eta_a\le0.995$, for the parameters of Eq.~\eqref{eq:params}. The coherence maximum occurs on the displayed $\eta_{\rm R}=0.995$ boundary.}
\label{tab:sizeopt}
\begin{ruledtabular}
\setlength{\tabcolsep}{2pt}\begin{tabular}{cccc}
$N_a$ & $C_{\rm mem}$ maximum & $|J_N|$ maximum & $J_E$ maximum \\
14 & $(0.43534,0.99500)$ & $(0.51760,0.89860)$ & $(0.51188,0.99500)$ \\
28 & $(0.42529,0.99500)$ & $(0.51759,0.89861)$ & $(0.52424,0.98317)$ \\
48 & $(0.42448,0.99500)$ & $(0.51759,0.89861)$ & $(0.52418,0.98329)$ \\
64 & $(0.42460,0.99500)$ & $(0.51759,0.89861)$ & $(0.52418,0.98329)$
\end{tabular}
\end{ruledtabular}
\end{table}

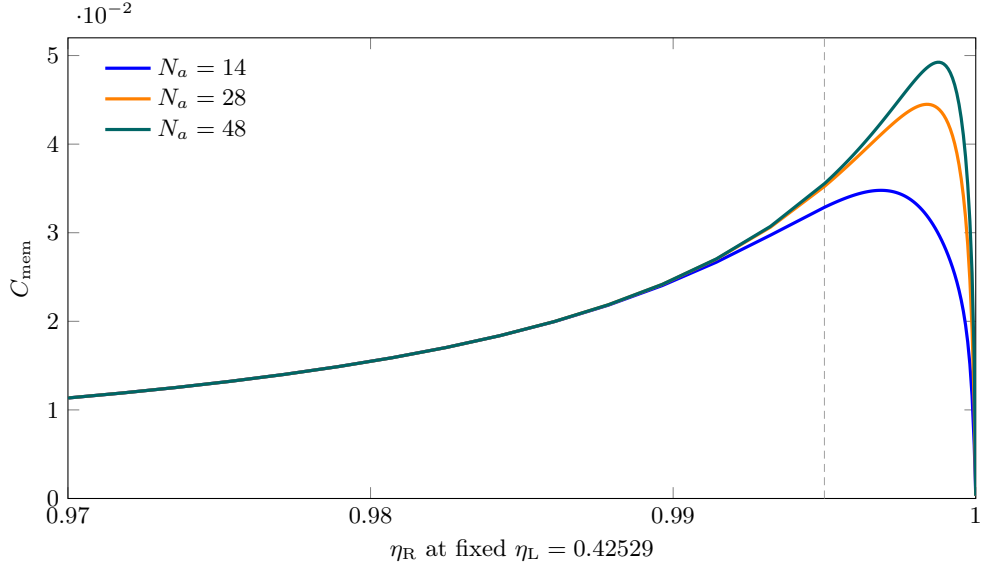
\begin{figure}[t]
\centering
\begin{tikzpicture}
\begin{axis}[width=0.78\columnwidth,height=0.52\columnwidth,scale only axis,
xmin=0.97,xmax=1,ymin=0,ymax=0.052,
xlabel={$\eta_{\rm R}$ at fixed $\eta_{\rm L}=0.42529$},ylabel={$C_{\rm mem}$},
xtick={0.97,0.98,0.99,1},
ytick={0,0.01,0.02,0.03,0.04,0.05},
grid=none,axis lines=box,
legend style={at={(0.03,0.98)},anchor=north west,draw=none,fill=white,font=\small},
tick label style={font=\small},label style={font=\small}]
\addplot[gray!70,densely dashed,thin,forget plot] coordinates {(0.995,0) (0.995,0.052)};
\addplot[very thick,blue] table[row sep=crcr] {
0.97 0.01134506539458\\
0.971785714 0.01190632238691\\
0.973571429 0.01252890750459\\
0.975357143 0.01322348534206\\
0.977142857 0.01400332291461\\
0.978928571 0.01488510583585\\
0.980714286 0.01589005466498\\
0.9825 0.01704542959121\\
0.984285714 0.01838649596947\\
0.986071429 0.0199588152008\\
0.987857143 0.02182003056919\\
0.989642857 0.02403798304944\\
0.991428571 0.02667447705127\\
0.993214286 0.02972009690263\\
0.995 0.03286827934404\\
0.995071739 0.03298434145639\\
0.995143478 0.03309863979706\\
0.995215217 0.03321102546171\\
0.995286957 0.03332134338821\\
0.995358696 0.03342942588458\\
0.995430435 0.03353510149666\\
0.995502174 0.03363818845025\\
0.995573913 0.0337384958152\\
0.995645652 0.03383582310388\\
0.995717391 0.03392995985669\\
0.99578913 0.0340206852148\\
0.99586087 0.03410776866798\\
0.995932609 0.0341909647975\\
0.996004348 0.03427002010151\\
0.996076087 0.03434466761133\\
0.996147826 0.03441462764799\\
0.996219565 0.03447960733046\\
0.996291304 0.03453930007664\\
0.996363043 0.03459338509906\\
0.996434783 0.03464152752467\\
0.996506522 0.03468337528328\\
0.996578261 0.03471856263285\\
0.99665 0.03474670688267\\
0.996721739 0.03476740860959\\
0.996793478 0.03478025117197\\
0.996865217 0.03478480023797\\
0.996936957 0.03478060321137\\
0.997008696 0.03476718915681\\
0.997080435 0.03474406805665\\
0.997152174 0.03471073054561\\
0.997223913 0.03466664758029\\
0.997295652 0.03461127014378\\
0.997367391 0.03454402899091\\
0.99743913 0.03446433443321\\
0.99751087 0.03437157477299\\
0.997582609 0.03426512150317\\
0.997654348 0.03414432142054\\
0.997726087 0.03400850144704\\
0.997797826 0.03385696721319\\
0.997869565 0.03368900272212\\
0.997941304 0.03350386980359\\
0.998013043 0.03330080722733\\
0.998084783 0.03307902606368\\
0.998156522 0.03283772010386\\
0.998228261 0.03257604396102\\
0.9983 0.03229312010834\\
0.998371739 0.03198802844216\\
0.998443478 0.03165979534194\\
0.998515217 0.03130737780115\\
0.998586957 0.0309296351027\\
0.998658696 0.03052531738094\\
0.998730435 0.03009298698768\\
0.998802174 0.02963097175522\\
0.998873913 0.02913726317174\\
0.998945652 0.02860937713457\\
0.999017391 0.02804415250106\\
0.99908913 0.02743745762136\\
0.99916087 0.02678374947592\\
0.999232609 0.02607546935034\\
0.999304348 0.02530202111061\\
0.999376087 0.0244483225556\\
0.999447826 0.02349246821441\\
0.999519565 0.0224019738672\\
0.999591304 0.02112749582729\\
0.999663043 0.01959179623341\\
0.999734783 0.01766889759684\\
0.999806522 0.01514035149673\\
0.999878261 0.01158402339882\\
0.99995 0.005987655979024\\
1 0.0003085060458495\\
};
\addlegendentry{$N_a=14$}
\addplot[very thick,orange] table[row sep=crcr] {
0.97 0.01134489998824\\
0.971785714 0.01190587516929\\
0.973571429 0.01252805272574\\
0.975357143 0.0132220800733\\
0.977142857 0.01400123691559\\
0.978928571 0.01488230231572\\
0.980714286 0.0158867914314\\
0.9825 0.01704274861276\\
0.984285714 0.01838741678373\\
0.986071429 0.01997128141707\\
0.987857143 0.02186431384235\\
0.989642857 0.02416565000752\\
0.991428571 0.02701787655356\\
0.993214286 0.03062302548664\\
0.995 0.0352264602225\\
0.995071739 0.03543482789032\\
0.995143478 0.03564498300837\\
0.995215217 0.0358569051483\\
0.995286957 0.03607057278594\\
0.995358696 0.03628595098037\\
0.995430435 0.03650300882685\\
0.995502174 0.03672170711163\\
0.995573913 0.03694200078045\\
0.995645652 0.03716383839573\\
0.995717391 0.0373871615423\\
0.99578913 0.03761190417867\\
0.99586087 0.03783799508729\\
0.995932609 0.03806534447193\\
0.996004348 0.038293862017\\
0.996076087 0.03852344336517\\
0.996147826 0.03875397223554\\
0.996219565 0.03898531929888\\
0.996291304 0.03921734094197\\
0.996363043 0.03944987790842\\
0.996434783 0.03968275705041\\
0.996506522 0.03991577669023\\
0.996578261 0.04014872428338\\
0.99665 0.04038136142848\\
0.996721739 0.04061342489308\\
0.996793478 0.04084462415909\\
0.996865217 0.0410746387029\\
0.996936957 0.04130311815102\\
0.997008696 0.04152966620947\\
0.997080435 0.04175385605243\\
0.997152174 0.04197521340128\\
0.997223913 0.04219321508407\\
0.997295652 0.04240728385107\\
0.997367391 0.04261678256574\\
0.99743913 0.04282100768289\\
0.99751087 0.04301918462505\\
0.997582609 0.04321044853239\\
0.997654348 0.04339385148435\\
0.997726087 0.04356834071612\\
0.997797826 0.04373274931075\\
0.997869565 0.04388578234826\\
0.997941304 0.04402600100186\\
0.998013043 0.04415180419008\\
0.998084783 0.04426140870171\\
0.998156522 0.04435281847633\\
0.998228261 0.04442380465363\\
0.9983 0.04447186470968\\
0.998371739 0.04449418295539\\
0.998443478 0.04448758101159\\
0.998515217 0.04444845760051\\
0.998586957 0.04437271320704\\
0.998658696 0.04425566272415\\
0.998730435 0.0440919131539\\
0.998802174 0.04387521884948\\
0.998873913 0.04359828803085\\
0.998945652 0.04325252840925\\
0.999017391 0.04282770539122\\
0.99908913 0.04231147405925\\
0.99916087 0.04168871574407\\
0.999232609 0.0409406405253\\
0.999304348 0.04004337118405\\
0.999376087 0.03896593124031\\
0.999447826 0.03766708473333\\
0.999519565 0.03609031872994\\
0.999591304 0.0341555463594\\
0.999663043 0.03174470050886\\
0.999734783 0.02867487084483\\
0.999806522 0.02464251585663\\
0.999878261 0.01907747867685\\
0.99995 0.01052924458171\\
1 0.0003084957883292\\
};
\addlegendentry{$N_a=28$}
\addplot[very thick,teal!80!black] table[row sep=crcr] {
0.97 0.01134491105929\\
0.971785714 0.0119058949677\\
0.973571429 0.01252808743772\\
0.975357143 0.01322213997206\\
0.977142857 0.01400133910686\\
0.978928571 0.01488247578105\\
0.980714286 0.01588708728831\\
0.9825 0.01704326365773\\
0.984285714 0.01838835409365\\
0.986071429 0.01997312175142\\
0.987857143 0.02186833806529\\
0.989642857 0.0241756399474\\
0.991428571 0.02704604443897\\
0.993214286 0.03071235096248\\
0.995 0.03554213028261\\
0.995071739 0.03576763100321\\
0.995143478 0.0359959040731\\
0.995215217 0.03622698797551\\
0.995286957 0.03646092396799\\
0.995358696 0.03669774282614\\
0.995430435 0.03693748421075\\
0.995502174 0.03718018343932\\
0.995573913 0.03742587447011\\
0.995645652 0.03767458960961\\
0.995717391 0.03792635917894\\
0.99578913 0.03818121113428\\
0.99586087 0.03843917425163\\
0.995932609 0.0387002632077\\
0.996004348 0.03896449959282\\
0.996076087 0.03923189693177\\
0.996147826 0.03950246352891\\
0.996219565 0.03977620163661\\
0.996291304 0.04005310650704\\
0.996363043 0.04033316530979\\
0.996434783 0.04061635986488\\
0.996506522 0.04090264939673\\
0.996578261 0.04119199260448\\
0.99665 0.04148433006742\\
0.996721739 0.04177958605419\\
0.996793478 0.04207766609717\\
0.996865217 0.04237845421014\\
0.996936957 0.04268181393923\\
0.997008696 0.04298756774067\\
0.997080435 0.04329551813239\\
0.997152174 0.0436054259495\\
0.997223913 0.04391700894878\\
0.997295652 0.044229935339\\
0.997367391 0.0445438163066\\
0.99743913 0.04485819737068\\
0.99751087 0.04517255275165\\
0.997582609 0.04548625623635\\
0.997654348 0.0457985940112\\
0.997726087 0.04610873146873\\
0.997797826 0.04641569937204\\
0.997869565 0.04671837262504\\
0.997941304 0.04701544550068\\
0.998013043 0.04730540269512\\
0.998084783 0.04758648929985\\
0.998156522 0.04785665553849\\
0.998228261 0.04811353369204\\
0.9983 0.04835436741476\\
0.998371739 0.04857595103201\\
0.998443478 0.04877455306291\\
0.998515217 0.04894582552183\\
0.998586957 0.04908469762658\\
0.998658696 0.04918523933229\\
0.998730435 0.04924051766076\\
0.998802174 0.04924239999038\\
0.998873913 0.04918132553072\\
0.998945652 0.0490460193151\\
0.999017391 0.04882313192415\\
0.99908913 0.04849677640035\\
0.99916087 0.04804791064136\\
0.999232609 0.04745353435412\\
0.999304348 0.04668548203594\\
0.999376087 0.0457087239215\\
0.999447826 0.04447869806398\\
0.999519565 0.04293695477738\\
0.999591304 0.04100354560276\\
0.999663043 0.03856259637913\\
0.999734783 0.03543177764831\\
0.999806522 0.03128707232536\\
0.999878261 0.02542291504911\\
0.99995 0.01556543529959\\
1 0.0003084957891273\\
};
\addlegendentry{$N_a=48$}
\end{axis}\end{tikzpicture}
\caption{High-retention behavior of the stationary coherence $C_{\rm mem}$ for several reservoir lengths $N_{\rm L}=N_{\rm R}=N_a$, at fixed $\eta_{\rm L}=0.42529$. The coherence has a narrow maximum within $0.97\le\eta_{\rm R}\le1$ that shifts toward full retention as $N_a$ increases. This maximum lies outside the range $\eta_{\rm R}\le0.995$ used for the operating landscape in Fig.~\ref{fig:maps}.}
\label{fig:edge}
\end{figure}

\paragraph{High-retention boundary layer beyond the displayed operating window.} The operating-window boundary $\eta_{\rm R}=0.995$ does not resolve the full high-retention region. At fixed $\eta_{\rm L}=0.42529$, the exact physical $N=28$ coherence increases from $C_{\rm mem}=0.03523$ at $\eta_{\rm R}=0.995$ to $0.04294$ at $\eta_{\rm R}=0.999$, and reaches an even larger local line maximum $0.04450$ near $\eta_{\rm R}=0.99839$. The window-constrained value $C_{\rm mem}=0.03523$ is therefore \emph{not} the maximum over the full physical retention range. Figure~\ref{fig:edge} resolves the thin high-retention region directly. With increasing chain length, the location of its conditional peak drifts toward unity: for $N=14,28,48,64$ the peak along the fixed-$\eta_{\rm L}$ cut is located near $0.99687,0.99839,0.99877,0.99896$, with coherence $0.03478,0.04450,0.04925,0.05142$, respectively. These are numerically optimized \emph{one-dimensional} cut maxima, not rigorous global extrema. For these four peak states the Stein residual is below $10^{-12}$ and the SPDM eigenvalues lie within $[0,1]$. At exactly $\eta_{\rm R}=1$ the current vanishes by the terminal-blockade lemma, while the memory on this particular cut falls sharply in the unique stationary solutions tested. This sharpened finite-size boundary-layer behavior makes extrapolation to $N=\infty$ especially delicate and prevents a universal monotonic-memory claim.

\section{Hard-block replacement and its physicality bound}\label{sec:hard}
For comparison, consider an algebraic hard-block prescription in which bath populations are reset exactly while a chosen fraction of system--bath coherence is retained. The post-intervention SPDM is postulated to have the form
\begin{equation}\widetilde C^+=\begin{pmatrix}P&x_{\rm L}&x_{\rm R}\\x_{\rm L}^\dagger&F_{\rm L}&0\\x_{\rm R}^\dagger&0&F_{\rm R}\end{pmatrix},\label{eq:hard}\end{equation}
where $x_a$ is a fraction $\eta_a$ of the pre-reset coherence. This generates an exact \emph{algebraic} affine map for $P,x_{\rm L},x_{\rm R}$, but exact linear algebra does not establish that the output is an allowed SPDM. With all target occupations strictly between zero and unity, applying the Schur complement to both $\widetilde C^+$ and $\Id-\widetilde C^+$ gives the \emph{necessary and sufficient} physicality conditions
\begin{align}
P&\ge\sum_{a,k}\frac{|x_{ak}|^2}{f_{ak}},\label{eq:hardlower}\\
1-P&\ge\sum_{a,k}\frac{|x_{ak}|^2}{1-f_{ak}}.\label{eq:hardupper}
\end{align}
The second constraint is the Pauli upper bound; checking positivity alone is insufficient. In fact, for fixed diagonal targets $0<F<I$, the formal overwrite is covariance-admissibility preserving for \emph{all} physical inputs if and only if
\begin{equation}
\eta_a^2\le \min_k\{f_{ak},1-f_{ak}\}\quad\text{for each }a.
\label{eq:uniformhard}
\end{equation}
To see sufficiency, let $y_i$ be the incoming system--bath coherence. From $C-C^2\ge0$, $\sum_i|y_i|^2\le P(1-P)$. Under Eq.~\eqref{eq:uniformhard}, both $\sum_i\eta_i^2|y_i|^2/f_i$ and $\sum_i\eta_i^2|y_i|^2/(1-f_i)$ are at most $P(1-P)$, which is no larger than either $P$ or $1-P$. For necessity choose a physical rank-one projector on the system and one bath mode, with $|y_i|^2=P(1-P)$ and all other cross elements zero. The lower Schur bound implies $\eta_i^2(1-P)\le f_i$ for every $0<P<1$; letting $P\downarrow0$ gives $\eta_i^2\le f_i$. The upper Schur bound gives $\eta_i^2 P\le1-f_i$; letting $P\uparrow1$ gives the other constraint. This result concerns covariance admissibility, not complete positivity of a state-independent many-body channel. Both conditions may fail at large coherence retention because very weakly occupied (or nearly filled) modes tightly bound their allowed cross-correlations.

\begin{figure}[t]
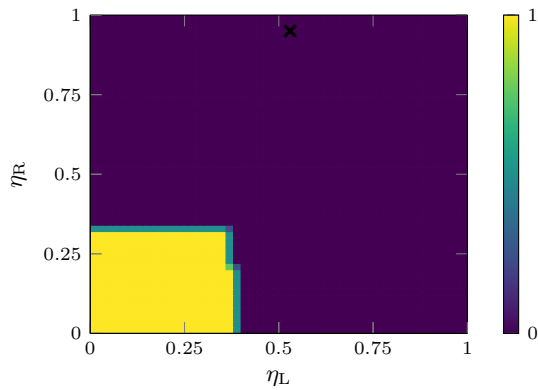
\centering

\caption{Fermionic admissibility domain of the \emph{algebraic hard-block} prescription. Yellow indicates post-reset fixed points satisfying \emph{both} Schur-complement inequalities \eqref{eq:hardlower} and \eqref{eq:hardupper}; dark indicates their violation. The cross marks a representative algebraic operating point that fails the fermionic admissibility test. The figure is computed from the corresponding affine algebraic map using $N_{\rm L}=N_{\rm R}=28$.}\label{fig:hard}
\end{figure}

At a representative algebraic hard-block operating point $(\eta_{\rm L},\eta_{\rm R})=(0.53,0.95)$, the same benchmark yields a negative smallest eigenvalue of $\widetilde C_\fp$ (approximately $-2.19\times10^{-2}$) and a largest eigenvalue approximately $1.0001$. The point therefore \emph{cannot} be interpreted as a fermionic stationary state. Figure~\ref{fig:hard} maps the Schur-complement admissibility test for this algebraic prescription; only part of the displayed parameter domain corresponds to a physical fermionic covariance. On the sampled $51\times51$ grid, the unrestricted algebraic maxima of the three observables lie outside the admissible region. Restricting the scan to its admissible points yields the same grid maximizer $(0.36,0.32)$ for all three diagnostics in this benchmark; no claim about continuous constrained maxima is made. The affine equations are algebraically well defined, and some solutions satisfy the covariance bounds. A physical interpretation nevertheless requires a state-by-state admissibility test and an independently specified implementation. A more general \emph{joint} intervention might retain correlations while fixing selected marginals, but its existence for a specified retention rule must be established separately. The algebraic hard-block results therefore do not, by themselves, establish the behavior of a physically implementable reset channel.

A bath-only \emph{replacement channel} mapping any bath state to its full thermal state also destroys all system--bath correlations: its bipartite action is $\rho_{S a}\mapsto\rho_S\otimes\rho_a^{\rm th}$. Thus a universal exact replacement of the full bath marginal combined with nonzero retained cross-correlations cannot be realized by a bath-only replacement channel. A more general protocol (for example one with joint access or restricted inputs) needs a separate physicality proof. The Gaussian attenuator explicitly resolves this issue by making bath renewal partial whenever coherence survives.

\section{Discussion and conclusions}\label{sec:disc}
The distinction between preserving correlations and restoring a thermal environment is central to the present two-reservoir protocol. Exact quadratic propagation does not guarantee that an arbitrary affine modification of the SPDM is a physical state transformation. The microscopic attenuator respects both fermionic inequalities, whereas the formal overwrite is admissible for all covariances only under the sharp threshold derived in Sec.~\ref{sec:hard}. On each side coherence retention and thermal renewal are linked by the channel, even though the left and right attenuation amplitudes can be independently selected. The Pauli bound limits the stored one-body coherence, and Eqs.~\eqref{eq:fidpresbound} and \eqref{eq:fidrefbound} relate the corresponding row norms to full-state distinguishability.

The full-state fidelities expose a distinct competition: an intervention that barely disturbs the state need not restore it toward the thermal product target. The high-retention cut exhibits a narrow coherence maximum beyond the original scan window, accompanied by a sizable reference-fidelity deficit. The reference fidelity compares the entire retained state with a product target; it is not the fidelity of a bath marginal alone. Its peak shifts with reservoir size, so the finite-grid optimum cannot establish an unrestricted optimum even at fixed size. The limits of long stroboscopic time, large reservoir size, and vanishing refresh can be nonuniform. Fidelity itself depends on the size of the compared states. Neither the stored coherence nor either fidelity establishes a quantum-memory channel capacity.

The exact switching-work ledger and correlation-inclusive entropy identity distinguish energy exchange from thermodynamic irreversibility. Reciprocity of the mode kernel gives a pairwise nonnegative entropy representation without assuming elastic transfer. Even two identical reservoirs with symmetric renewal exhibit positive switching work and nontrivial coherence and fidelity, although their net particle transfers vanish by symmetry. The biased benchmark likewise shows that zero particle throughput does not imply zero switching work: at $(0.25,1)$, $J_N=0$ while $J_E\simeq0.00714$. The continuity and short-cycle transfer laws delimit the particle-exchange contribution to the broader reset problem; the operating-point separation is not a universal transport-optimization theorem. Maximizing the signed $J_E$ does not minimize a thermodynamic resource cost.

The idealized channel requires access to the refreshed lead. For uniform attenuation, identical rotations between corresponding sites of two identical chains implement the same one-particle transformation, because the rotation commutes with their common standing-wave basis change. Endpoint-only contact is a different operation. Work invested in preparing and disposing of fresh ancillas and in nonideal contact control is not included in the switching-work ledger. Extensions to unequal bandwidths, thermal bias, disorder, finite-duration contacts, and resource-constrained fidelity optimization may clarify how the distinction between preserving correlations and restoring a target persists in more general settings. The present exact finite-period Gaussian framework provides a benchmark for these questions.

\begin{acknowledgments}
J.K. and T.A.-N. have been supported by the Academy of Finland through its QTF Center of Excellence (grant No.~312298), by the European Union and the European Innovation Council through the Horizon Europe project QRC-4-ESP (grant No.~101129663), and by the EU Horizon Europe Quest project (No.~10116088). A.L. acknowledges support from the Leverhulme Trust Research Project Grant RPG-2025-063.
\end{acknowledgments}

\section*{Data availability}
The numerical data for all figures are embedded in the manuscript source file. The Python codes to reconstruct the finite-chain Hamiltonians, solve the stationary covariance equations, and reproduce the plotted values and consistency tests are available from the authors upon reasonable request.

\section*{Conflicts of interest}
The authors declare no conflicts of interest.

\appendix

\section{Complete two-reservoir block propagation}\label{app:fullblocks}
This appendix gives the complete SPDM block algebra for the Gaussian partial-refresh operation, including the evolution of \emph{all} inter-bath correlations. No bath block is assumed diagonal at the beginning of a general physical steady-state cycle. We use $a,b,c\in\{\mathrm L,\mathrm R\}$, reserving $k$ for a mode index within a given bath.

\subsection{One-cycle unitary block multiplication}
At the start of the $n$th unitary segment write the full Hermitian covariance as
\begin{equation}
C_n^+=\begin{pmatrix}
P_n&x_{{\rm L},n}&x_{{\rm R},n}\\
x_{{\rm L},n}^\dagger&B_{{\rm LL},n}&B_{{\rm LR},n}\\
x_{{\rm R},n}^\dagger&B_{{\rm RL},n}&B_{{\rm RR},n}
\end{pmatrix},\quad B_{ba,n}=B_{ab,n}^\dagger.
\label{eq:fullC}
\end{equation}
The blocks of the exact one-particle propagator are
\begin{equation}
U=\begin{pmatrix}u&v_{\rm L}&v_{\rm R}\\
w_{\rm L}&X_{\rm LL}&X_{\rm LR}\\
w_{\rm R}&X_{\rm RL}&X_{\rm RR}\end{pmatrix}.
\label{eq:Ublocks}
\end{equation}
Here $u$ is a scalar, $v_a$ are rows, $w_a$ are columns, and $X_{ab}$ maps bath $b$ to bath $a$. Even though $M_{\rm LR}=0$, generally $X_{\rm LR}\ne0$ for nonzero cycle duration: excursions between the baths are mediated by the system level. The exact statement $U^\dagger U=\Id$ supplies important numerical checks.

First multiply the matrices $T_n=UC_n^+$. Their system--system and system--bath blocks are
\begin{align}
T_{00}&=uP_n+\sum_c v_c x_{c,n}^\dagger,\label{eq:T00}\\
T_{0b}&=ux_{b,n}+\sum_c v_cB_{cb,n}.
\label{eq:T0b}
\end{align}
The corresponding bath-row blocks follow from the second and third rows of $U$:
\begin{align}
T_{a0}&=w_aP_n+\sum_c X_{ac}x_{c,n}^\dagger,\label{eq:Ta0}\\
T_{ab}&=w_ax_{b,n}+\sum_c X_{ac}B_{cb,n}.
\label{eq:Tab}
\end{align}
A second multiplication by $U^\dagger$ gives all independent blocks of the pre-refresh covariance $C_n^-=T_nU^\dagger$:
\begin{align}
P_n^-&=T_{00}u^*+\sum_bT_{0b}v_b^\dagger,\label{eq:Pminusblock}\\
x_{a,n}^-&=T_{00}w_a^\dagger+\sum_bT_{0b}X_{ab}^\dagger,
\label{eq:xminusblock}\\
B_{ab,n}^-&=T_{a0}w_b^\dagger+\sum_cT_{ac}X_{bc}^\dagger.
\label{eq:Bminusblock}
\end{align}
These three equations together with Eqs.~\eqref{eq:T00}--\eqref{eq:Tab} explicitly evaluate every component of $UC_n^+U^\dagger$. For example, the system occupation can be expanded into a system-population term, system--bath cross terms, and bath-block terms:
\begin{align}
P_n^-={}&|u|^2P_n
+\sum_a\bigl(u^*v_ax_{a,n}^\dagger
+ux_{a,n}v_a^\dagger\bigr)
\nonumber\\
&+\sum_{a,b}v_aB_{ab,n}v_b^\dagger.
\label{eq:PminusExpanded}
\end{align}
For the left bath the pre-refresh block is, without suppressing either cross-bath coherence,
\begin{align}
B_{{\rm LL},n}^-={}&T_{{\rm L}0}w_{\rm L}^\dagger
+T_{{\rm LL}}X_{{\rm LL}}^\dagger
+T_{{\rm LR}}X_{{\rm LR}}^\dagger,
\label{eq:BLminusExpanded}
\end{align}
while its right-bath counterpart is
\begin{align}
B_{{\rm RR},n}^-={}&T_{{\rm R}0}w_{\rm R}^\dagger
+T_{{\rm RL}}X_{{\rm RL}}^\dagger
+T_{{\rm RR}}X_{{\rm RR}}^\dagger.
\label{eq:BRminusExpanded}
\end{align}
The mixed block $B_{\rm LR}^-$ is generated in the same multiplication and must also be carried into the next cycle; setting it to zero would define a different operation.

\subsection{Physical refresh of the full block matrix}
The ancilla contact acts on those pre-refresh blocks as
\begin{align}
P_{n+1}&=P_n^-,\label{eq:fullPmap}\\
x_{a,n+1}&=\eta_a x_{a,n}^-,\label{eq:fullxmap}\\
B_{ab,n+1}&=\eta_a\eta_b B_{ab,n}^-
+\delta_{ab}(1-\eta_a^2)F_a.
\label{eq:fullBmap}
\end{align}
In particular,
\begin{equation}
B_{{\rm LR},n+1}=\eta_{\rm L}\eta_{\rm R}
B_{{\rm LR},n}^-.
\label{eq:fullLR}
\end{equation}
Equations~\eqref{eq:T00}--\eqref{eq:fullLR} define a closed affine map on the full Hermitian covariance matrix. They reduce identically to Eq.~\eqref{eq:affine} after setting $A=KU$. They also show why a reduced map retaining only the system occupation and two system--bath coherences does not close for the partial-refresh operation: the physical map contains dynamically evolving $B_{aa}$ and $B_{\rm LR}$ blocks even at stationarity.

The pre-contact occupation and energy transfers may be recovered without reconstructing the outgoing ancilla covariance explicitly:
\begin{align}
\Delta N_a&=\Tr(B_{aa}^--B_{aa}),\label{eq:fullDN}\\
Q_a&=\Tr[M_a(B_{aa}^--B_{aa})].
\label{eq:fullQ}
\end{align}
Combining Eq.~\eqref{eq:fullBmap} with these expressions gives Eqs.~\eqref{eq:dN} and \eqref{eq:energy} directly, with the partial-renewal factors. In particular, the derivation retains the actual bath block $B_{aa}$, which generally differs from $F_a$ when $\eta_a>0$.

\subsection{Fixed-point uniqueness and conditioning}
Column-wise vectorization uses $\operatorname{vec}(AC A^\dagger)=(A^*\otimes A)\operatorname{vec}C$. Thus the full affine map has linear coefficient $A^*\otimes A$ on vectorized covariances and inhomogeneity $\operatorname{vec}D$. Its spectral radius is $r(A)^2$. If $r(A)<1$, then $\Id-A^*\otimes A$ is invertible and the Neumann series converges in finite dimension, establishing Eq.~\eqref{eq:series}. A useful necessary-and-sufficient stability test follows without any perturbation expansion. Since $K^\dagger K=K^2\le\Id$, for an eigenvector $Av=\lambda v$ with $|\lambda|=1$ we have
\begin{equation}
\lVert v\rVert=\lVert KUv\rVert\le\lVert Uv\rVert=\lVert v\rVert.
\end{equation}
Equality implies $(\Id-K^2)Uv=0$; therefore $KUv=Uv$ and $Uv=\lambda v$. In particular, $v\in\ker(\Id-K^2)$. Conversely, any eigenvector of $U$ in this undamped subspace is also a unit-modulus eigenvector of $A=KU$. Hence $r(A)<1$ \emph{if and only if} $U$ has no eigenvector fully contained in $\ker(\Id-K^2)$. This test includes the one-sided no-refresh edges and avoids treating the simpler two-sided damped criterion as universal.

Close to a nonrefreshing endpoint, however, $r(A)$ may approach one, making the inverse poorly conditioned and steady-state preparation slow. Indeed for $N_{\rm L}=N_{\rm R}=28$ and the stated Hamiltonian, direct diagonalization gives $r(KU)\simeq0.99999999025$ on either $(\eta_{\rm L},\eta_{\rm R})=(1,0.25)$ or $(0.25,1)$. Although strictly inside the unit circle at the reported precision, this corresponds to a covariance relaxation eigenvalue of modulus $r(KU)^2$ extraordinarily close to unity and a characteristic worst-mode scale $[1-r(KU)^2]^{-1}\sim5\times10^7$ cycles. The number is a spectral slow-mode timescale, not necessarily the observed equilibration time of every initial state or observable. Therefore numerical residuals, conditioning, and finite-size behavior are distinct checks; a small algebraic residual does not by itself prove that a high-retention state is reached quickly in a laboratory.

\section{Gaussian ancilla dilation and complete positivity}\label{app:cp}
For each bath mode, form the column $(\hat a_{ak},\hat b_{ak})^T$ and the orthogonal, particle-number-conserving mixing matrix
\begin{equation}B_a=\begin{pmatrix}\eta_a&\sqrt{1-\eta_a^2}\\-\sqrt{1-\eta_a^2}&\eta_a\end{pmatrix},\qquad B_a^\dagger B_a=\Id.\end{equation}
The incoming ancillas have thermal covariance $F_a$, have no anomalous pair expectations, and are initially uncorrelated with the retained system and baths. Expanding products of the transformed operators in Eq.~\eqref{eq:rotate} and setting all initial system--ancilla cross correlators to zero gives, explicitly,
\begin{align}
\langle (\hat a_{a\ell}^+)^\dagger\hat a_{ak}^+\rangle
 &=\eta_a^2(C_{aa}^-)_{k\ell}+(1-\eta_a^2)(F_a)_{k\ell},\\
\langle (\hat a_{ak}^+)^\dagger\hat a_0^+\rangle
 &=\eta_a(C_{0a}^-)_{k},\nonumber\\
\langle (\hat a_{{\rm R}q}^+)^\dagger\hat a_{{\rm L}k}^+\rangle
 &=\eta_{\rm L}\eta_{\rm R}(C_{\rm L R}^-)_{kq}.\nonumber
\end{align}
The system occupation is unchanged. Collecting these entries proves Eq.~\eqref{eq:refresh}. This transformation is a restriction of a fermionic number-conserving unitary and is therefore a completely positive, trace-preserving Gaussian channel at the many-body level.

Because $D\ge0$ and $(\Id-K^2)-D\ge0$, Eq.~\eqref{eq:positivity} follows immediately. For a fixed point, Eq.~\eqref{eq:series} proves $C_\fp\ge0$. Further, the hole matrix $H=\Id-C$ obeys $H_{n+1}=A H_n A^\dagger+[(\Id-K^2)-D]$. Its stationary solution is likewise positive, giving $C_\fp\le\Id$. This second inequality is indispensable for fermions. More precisely, if $r(A)<1$ and all $f_i$ are strictly between zero and one, set $f_{\min}=\min_i f_i$ and $f_{\max}=\max_i f_i$. Since $D=LFL^\dagger$ and $LL^\dagger=I-AA^\dagger$, telescoping the stationary series gives
\begin{equation}
f_{\min}I\le C_\fp\le f_{\max}I.
\label{eq:strictinterior}
\end{equation}
Indeed $f_{\min}LL^\dagger\le D\le f_{\max}LL^\dagger$, and $\sum_{m\ge0}A^mLL^\dagger(A^\dagger)^m=I$. The same bounds hold for $C_\fp^-$ by unitary conjugation. This proves that the many-body Gaussian density operators in the finite-temperature stable benchmark have full rank, including a singly unrefreshed edge whenever it is stable.

If $A\psi=\lambda\psi$ with $|\lambda|=1$ and $U$ unitary, then
\begin{equation}\|\psi\|=\|K U\psi\|\le\|U\psi\|=\|\psi\|.\end{equation}
For both $\eta_a<1$, equality demands that $U\psi$ have zero projection on both baths. Since $A\psi=\lambda\psi$ this also demands that $\psi$ be supported exclusively on the system. Hence absence of a system-only eigenvector of $U$ rules out any unit-modulus eigenvalue, proving $r(A)<1$. Exceptional revival times must be excluded when claiming uniqueness.

\section{Explicit derivation of the short-cycle kernel}\label{app:small}
Let $\delta_{ak}=P-f_{ak}$, $g=g_{ak}$, and $\Delta_{ak}=\omega_0-\omega_{ak}$. We first justify the orders used in the expansion. The admissibility bound $0\le C\le I$ makes the Taylor remainder of $U C U^\dagger$ uniformly bounded in $C$ at fixed $M$. For a stationary covariance, $x_{ak}=\eta_a (x_{ak}+O(\tau))$, which yields $x_{ak}=O(\tau)$ because $1-\eta_a>0$. For bath indices $i,j$ with attenuations $\eta_i,\eta_j$, stationarity then implies
\begin{equation}
(1-\eta_i\eta_j)(B_{ij}-f_i\delta_{ij})
=\eta_i\eta_j\bigl[-i\tau[M,C]_{ij}+O(\tau^2)\bigr].
\end{equation}
This first gives $B-F=O(\tau)$. Since $[M,C]_{ij}=(\omega_i-\omega_j)B_{ij}+g_i^*x_j-x_i^*g_j$ and $[M_E,F]=0$, the commutator is itself $O(\tau)$. Substituting back proves $B-F=O(\tau^2)$. The constants are finite at fixed attenuations below unity. For sufficiently small nonzero $\tau$ and at least one nonzero coupling, $U|0\rangle=|0\rangle-i\tau M|0\rangle+O(\tau^2)$ has a nonzero bath component; hence the stable stationary state exists by the criterion of Appendix~\ref{app:fullblocks}. On the stationary/relaxed manifold, the commutator in the coherence sector is
\begin{align}
[M,C]_{0,ak}&=\omega_0 x_{ak}+g f_{ak}-P g-x_{ak}\omega_{ak}+O(\tau^2)\nonumber\\
&=\Delta_{ak}x_{ak}-g\delta_{ak}+O(\tau^2).\label{eq:comx}
\end{align}
Using $C^-=C-i\tau[M,C]+O(\tau^2)$ and multiplying the refreshed coherence by $\eta_a$ gives
\begin{equation}(1-\eta_a)x_{ak}=i\eta_a g\delta_{ak}\tau+O(\tau^2),\end{equation}
which is Eq.~\eqref{eq:xasym}. The occupation change begins at second order after the stationary coherence terms are included. One must use
\begin{equation}P^- -P=-i\tau[M,C]_{00}-\frac{\tau^2}{2}[M,[M,C]]_{00}+O(\tau^3).\label{eq:double}
\end{equation}
The first commutator is
\begin{equation}[M,C]_{00}=\sum_{a,k}(g_{ak}x_{ak}^*-x_{ak}g_{ak}^*).\label{eq:c00}\end{equation}
Since $x_{ak}=i\eta_a g_{ak}\delta_{ak}\tau/(1-\eta_a)+O(\tau^2)$,
\begin{equation}-i\tau[M,C]_{00}=-2\tau^2\sum_{a,k}\frac{\eta_a}{1-\eta_a}|g_{ak}|^2\delta_{ak}+O(\tau^3).\label{eq:cohterm}\end{equation}
At zeroth order in $\tau$, $C=\diag(P,F_{\rm L},F_{\rm R})+O(\tau)$. For each bath mode,
\begin{align}
[M,C]_{0,ak}&=-g_{ak}\delta_{ak}+O(\tau),\\
[M,C]_{ak,0}&=g_{ak}^*\delta_{ak}+O(\tau).
\end{align}
Consequently, direct expansion of the nested commutator gives
\begin{align}
[M,[M,C]]_{00}
&=\sum_{a,k}g_{ak}[M,C]_{ak,0}\nonumber\\
&\quad-\sum_{a,k}[M,C]_{0,ak}g_{ak}^*\nonumber\\
&=2\sum_{a,k}|g_{ak}|^2\delta_{ak}+O(\tau).
\label{eq:dcomm}\end{align}
Thus the $-\tau^2[M,[M,C]]_{00}/2$ term equals $-\tau^2\sum_{a,k}|g_{ak}|^2\delta_{ak}+O(\tau^3)$. Adding both terms, the coefficient is
\begin{equation}1+\frac{2\eta_a}{1-\eta_a}=\frac{1+\eta_a}{1-\eta_a},\end{equation}
which proves Eq.~\eqref{eq:deltaP}. As the refresh leaves the system occupation unchanged, stationarity imposes $P^-=P$, giving Eq.~\eqref{eq:Pcondition}. For completeness the bath-mode occupation expansion can be performed independently. Write $b_{ak}=(C^+)_{ak,ak}=f_{ak}+O(\tau^2)$. With the same slaved coherence, the first commutator satisfies
\begin{align}
[M,C]_{ak,ak}&=g_{ak}^*x_{ak}-x_{ak}^*g_{ak}\\
&=\frac{2i\eta_a\tau}{1-\eta_a}|g_{ak}|^2\delta_{ak}
+O(\tau^2).
\label{eq:modecomm}
\end{align}
At zeroth order in the covariance correction the nested commutator is
\begin{align}
[M,[M,C]]_{ak,ak}
&=g_{ak}^*[M,C]_{0,ak}
-[M,C]_{ak,0}g_{ak}\\
&=-2|g_{ak}|^2\delta_{ak}+O(\tau).
\label{eq:modedcomm}
\end{align}
The second-order unitary expansion now yields
\begin{align}
\Delta b_{ak}&:=(C^-)_{ak,ak}-b_{ak},\nonumber\\
\Delta b_{ak}&=-i\tau[M,C]_{ak,ak}
-\frac{\tau^2}{2}[M,[M,C]]_{ak,ak}
+O(\tau^3)\nonumber\\
&=\tau^2|g_{ak}|^2
\frac{1+\eta_a}{1-\eta_a}\,\delta_{ak}
+O(\tau^3)\nonumber\\
&=\gamma_{ak}(\eta_a,\tau)\delta_{ak}+O(\tau^3).
\label{eq:modetransfer}
\end{align}
The incoming and outgoing ancillas are mode matched. Therefore, summing the change in each bath-mode occupation gives
$\Delta N_a=\sum_k[(C^-)_{ak,ak}-b_{ak}]$;
weighting each mode by its energy instead gives
$Q_a=\sum_k\omega_{ak}[(C^-)_{ak,ak}-b_{ak}]$.
These are exactly Eqs.~\eqref{eq:currentasym} and \eqref{eq:Qasym}, including the contribution from intra-cycle transitions at $\eta_a=0$. Bath--bath off-diagonal covariances first affect the displayed mode-resolved diagonal transfers beyond the order kept here for fixed $\eta_a<1$. These expansions have bounds uniform only on compact subsets of the domain $\eta_a<1$ at fixed finite Hilbert-space dimension; in particular, their constants can depend on bath size and spectral gaps. They must not be extrapolated to the undamped endpoint or used without a separate estimate for an infinite reservoir. In particular, setting $\eta_a=0$ before expanding leaves precisely the direct $|g_{ak}|^2\tau^2$ contribution.

\subsection{Particle-exchange coefficient as a consistency diagnostic}
For nonzero couplings to both baths define $G_a=\sum_k |g_{ak}|^2(1+\eta_a)/(1-\eta_a)$ and $\bar f_a=\sum_k |g_{ak}|^2 f_{ak}/\sum_k|g_{ak}|^2$. Then Eq.~\eqref{eq:Pweighted} gives $P_\fp=(G_{\rm L}\bar f_{\rm L}+G_{\rm R}\bar f_{\rm R})/(G_{\rm L}+G_{\rm R})+O(\tau)$. Inserting this in Eq.~\eqref{eq:currentasym} yields
\begin{equation}
J_N=\tau\frac{G_{\rm L}G_{\rm R}}{G_{\rm L}+G_{\rm R}}
(\bar f_{\rm L}-\bar f_{\rm R})+O(\tau^2).
\label{eq:seriescurrent}
\end{equation}
The positive series factor increases strictly with either $G_a$ because its derivative with respect to $G_a$ is $G_b^2/(G_a+G_b)^2$. Since $\bar f_a$ is independent of $\eta_a$, the magnitude of the leading current coefficient increases with either retention parameter for nonzero effective filling bias. This statement concerns the leading coefficient, not a uniform monotonicity theorem at finite $\tau$.

\section{Tight-binding spectral density}\label{app:spectrum}
For the finite open nearest-neighbor chain, the site-basis single-particle Hamiltonian is
\begin{equation}
M_a=-J_a\sum_{j=1}^{N_a-1}
\bigl(|j\rangle\langle j+1|+|j+1\rangle\langle j|\bigr).
\end{equation}
Its eigenproblem reads $-J_a(\phi_{j-1}+\phi_{j+1})=\omega\phi_j$ with open conditions $\phi_0=\phi_{N_a+1}=0$. The ansatz $\phi_j=\mathcal N\sin(kj)$ implies $\omega=-2J_a\cos k$ and quantization $k_m=m\pi/(N_a+1)$, $m=1,\ldots,N_a$. The normalization follows from
\begin{equation}
\sum_{j=1}^{N_a}\sin^2\!\left(\frac{m\pi j}{N_a+1}\right)=\frac{N_a+1}{2},
\end{equation}
so $\mathcal N=\sqrt{2/(N_a+1)}$. A contact hopping $t_a$ to site $j=1$ then couples to each eigenmode with $g_{am}=t_a\mathcal N\sin k_m$. These are the discrete data used in the exact numerical construction; replacing the sum by a continuum density is not done in the finite-size computation.

In the continuum limit $N_a\to\infty$, the open semi-infinite chain has delta-normalized standing-wave modes $\phi_j(k)=\sqrt{2/\pi}\sin(jk)$, $k\in(0,\pi)$, and $\omega(k)=-2J\cos k$. The surface contact amplitude is $g(k)=t\phi_1(k)$. By definition,
\begin{align}
\mathcal J(\omega)&=\int_0^\pi\!dk\,|g(k)|^2\delta[\omega+2J\cos k]\nonumber\\
&=\frac{2t^2}{\pi}\frac{\sin^2k_\omega}{2J\sin k_\omega}\nonumber\\
&=\frac{t^2}{2\pi J^2}\sqrt{4J^2-\omega^2},\qquad |\omega|<2J,
\end{align}
where $\cos k_\omega=-\omega/(2J)$ and $\sin k_\omega>0$. The density is zero outside the band. For a chain of length $N$, the exact discrete modes have $k_j=j\pi/(N+1)$ and $g_j=t\sqrt{2/(N+1)}\sin k_j$; these, rather than a discretized continuum integral, are used in all figures.

\section{Particle and energy ledger}\label{app:energy}
Write $H_0=\omega_0\hat n_0+\sum_a\hat a_a^\dagger M_a\hat a_a$ and $H_{\rm int}=\sum_a(\hat a_0^\dagger g_a\hat a_a+\mathrm{H.c.})$. The system occupation is unchanged by refresh, so
\begin{align}
\Delta E_{0,\rm refresh}&=\Tr[(\omega_0\oplus M_{\rm L}\oplus M_{\rm R})(C^+-C^-)]\nonumber\\
&=-\sum_a Q_a.\label{eq:E0change}
\end{align}
The one-body interaction-energy change is
\begin{equation}\Delta E_{\rm int,refresh}=2\operatorname{Re}\sum_{a,k}g_{ak}^*[(C^+)_{0,ak}-(C^-)_{0,ak}].\label{eq:Eint}
\end{equation}
At a stationary stroboscopic cycle, $\Tr M C^-=\Tr M C_\fp$ because $U$ commutes with $M$ and $C^+=C_\fp$, whence
\begin{equation}\sum_a Q_a=\Delta E_{\rm int,refresh}.\label{eq:ledger}
\end{equation}
The fresh ancillas gain energy $\sum_aQ_a$ and particle number $\sum_a\Delta N_a=0$. The local beam splitters conserve the bare bath-plus-ancilla energy and particle number because each exchanged mode pair has the same frequency. They need not conserve $H_{\rm int}$ when activated. In the sudden-coupling idealization, the external switching controller performs signed work equal to the full retained-plus-ancilla energy change, namely $W_{\rm sw}=\Delta E_{\rm int,refresh}=\sum_aQ_a$ over the stationary cycle, with the sign convention that positive $W_{\rm sw}$ means work supplied by the controller. The signed net work is not the sum of positive side-resolved heat magnitudes; if the protocol is implemented by another drive, its own work ledger must be derived separately.

For the ancillary modes on side $a$, $\beta_a(Q_a-\mu_a\Delta N_a)$ is their outgoing-minus-incoming grand-energy change multiplied by inverse temperature, not by itself their von Neumann entropy change. For the stationary \emph{Gaussian many-body state} corresponding to $C_{\fp}$, a rigorous entropy-production statement follows from the fresh-ancilla dilation: joint unitary evolution and entropy subadditivity yield
\begin{equation}\sum_a\beta_a(Q_a-\mu_a\Delta N_a)\ge0\label{eq:secondlaw}
\end{equation}
when all ancillas begin in independent Gibbs states and the retained Gaussian system--bath state returns to itself after the entire stroboscopic cycle. In the equal-temperature benchmark, $\beta_{\rm L}=\beta_{\rm R}=\beta$, and the exact continuity law makes this inequality equivalent to
\begin{equation}
\beta\tau\left[J_E-(\mu_{\rm R}-\mu_{\rm L})J_N\right]\geq0.
\label{eq:biasSecondLaw}
\end{equation}
This is a constraint on the combination of signed energy and particle fluxes; it does not force $J_E$ to be positive at arbitrary bias. In general, stationarity of an SPDM alone would not establish stationarity of higher-order correlations in an arbitrary non-Gaussian many-body state; here the Gaussian channel and its Gaussian fixed point justify the full-state entropy argument. The precise identity makes the statement stronger and identifies the physical entropy terms. Let $R$ denote \emph{all} retained system-plus-bath modes just before the ancilla contact, let $A$ denote all fresh ancilla modes, let $\omega_A=\bigotimes_a Z_a^{-1}e^{-\beta_a(H_{A,a}-\mu_a N_{A,a})}$, and let $\rho_{RA}^{\rm out}$ be the state immediately after their joint number-conserving contact. The incoming state is $\rho_R^-\otimes\omega_A$, while the retained output $\rho_R^+$ equals the state at the start of the unitary leg at stationarity. Since both the retained-system-plus-baths unitary leg and the joint contact conserve the respective total von Neumann entropies, $S(\rho_R^+)=S(\rho_R^-)$, and hence
\begin{align}
I(R{:}A)_{\rm out}&=S(\rho_A^{\rm out})-S(\omega_A),\nonumber\\
D(\rho_A^{\rm out}\Vert\omega_A)
 &=\sum_a\beta_a(Q_a-\mu_a\Delta N_a)\nonumber\\
 &\quad-[S(\rho_A^{\rm out})-S(\omega_A)].\label{eq:entropyid}
\end{align}
Adding these expressions gives the exact \emph{repeated-interaction entropy balance}
\begin{equation}
\sum_a\beta_a(Q_a-\mu_a\Delta N_a)=I(R{:}A)_{\rm out}+D(\rho_A^{\rm out}\Vert\omega_A)\geq0.\label{eq:entropyidentity}
\end{equation}
Here $I$ is mutual information and $D$ is quantum relative entropy. Esposito, Lindenberg, and Van den Broeck derived an entropy-production identity for initially thermal finite reservoirs \cite{EspositoLindenbergVdB2010}. Reeb and Wolf established an equality underlying the Landauer bound and derived finite-size corrections \cite{ReebWolf2014}. Deffner and Lutz studied nonequilibrium entropy production for open quantum systems \cite{DeffnerLutz2011}. Our stationary identity uses the full outgoing ancilla state and the grand-canonical energy changes of the individual contacts. This identity holds for the complete collection of outgoing ancillas; replacing its joint relative entropy by an arbitrary sum of bath-marginal relative entropies would in general discard correlations. No entropy-production identification with a single bath SPDM marginal is made. In the present model the distinction is especially relevant because the outgoing left- and right-side ancillas can become mutually correlated during their joint contact with the retained, initially correlated two-bath state. Relative entropy between the \emph{pre-refresh bath marginal} and $F_a$ alone is not, in general, the full entropy production of this correlated repeated-contact process. This information-theoretic accounting complements the entropy-production inequality developed by Spohn for quantum dynamical semigroups \cite{Spohn1978}.

\subsection{Finite-reservoir particle-exchange bounds}
Let $\nu_a=\Tr F_a$, with $0\le\nu_a\le N_a$. For any physical pre-contact covariance $0\le C^-\le\Id$, each bath-mode population lies between zero and one. Using Eq.~\eqref{eq:dN} mode by mode gives
\begin{equation}
-(1-\eta_a^2)\nu_a\le\Delta N_a\le(1-\eta_a^2)(N_a-\nu_a).
\label{eq:modebound}
\end{equation}
Applying $\Delta N_{\rm R}=\tau J_N=-\Delta N_{\rm L}$ and introducing $q_a=1-\eta_a^2$ gives the signed bounds
\begin{align}
-\min\{q_{\rm R}\nu_{\rm R},
 q_{\rm L}(N_{\rm L}-\nu_{\rm L})\}
&\le \tau J_N,\nonumber\\
\tau J_N&\le\min\{q_{\rm R}(N_{\rm R}-\nu_{\rm R}),
 q_{\rm L}\nu_{\rm L}\}.
\label{eq:tightcurrentbound}
\end{align}
These inequalities require neither a small-$\tau$ expansion nor knowledge of the stationary coherences. They establish an explicit upper bound on available particle throughput which collapses when either side has no renewal. They are finite-size constraints, and their right sides are generally not optimal bounds as $N_a\to\infty$.

\section{Hard-reset Schur complements and RI endpoint}\label{app:hard}
Let $F=F_{\rm L}\oplus F_{\rm R}$ and $x=(x_{\rm L},x_{\rm R})$. For $0<F<\Id$, the Schur complements of $\widetilde C^+=\begin{psmallmatrix}P&x\\x^\dagger&F\end{psmallmatrix}$ and $\Id-\widetilde C^+$ are, respectively, $P-xF^{-1}x^\dagger$ and $1-P-x(\Id-F)^{-1}x^\dagger$. This proves Eqs.~\eqref{eq:hardlower} and \eqref{eq:hardupper} as both necessary and sufficient. If some occupation reaches exactly zero or one, the corresponding pseudoinverse and range conditions must be imposed instead.

For complete coherence erasure, $x_a=0$ after each reset but is regenerated within the unitary segment. Starting from $\widetilde C^+=\diag(P,F)$,
\begin{align}P^-&=P-\tau^2\sum_{a,k}|g_{ak}|^2(P-f_{ak})+O(\tau^3),\nonumber\\
\Delta N_a&=\tau^2\sum_k |g_{ak}|^2(P-f_{ak})+O(\tau^3).
\end{align}
Thus the complete-erasure endpoint of the algebraic prescription also exhibits a nonzero leading transfer coefficient.

\section{Mode-transfer representation and continuous scaling}\label{app:kernel}
For the ancilla column $b$ the contact sends $b^{\rm out}=-L^\dagger a^-+H_\eta b$, where $a^-=Ua$ and $a$ is the retained annihilation column at the beginning of a period. Independence of fresh inputs gives the outgoing covariance
\begin{equation}
F^{\rm out}=L^\dagger U C_\fp U^\dagger L+H_\eta F H_\eta.
\end{equation}
Substitution of Eq.~\eqref{eq:series} proves Eq.~\eqref{eq:transferkernel} element by element. For row normalization note that $LL^\dagger=I-AA^\dagger$. Hence
\begin{align}
\sum_{m=0}^{p} A^m LL^\dagger(A^\dagger)^m
&=\sum_{m=0}^{p}[A^m(A^\dagger)^m\nonumber\\
&\hspace{5mm}-A^{m+1}(A^\dagger)^{m+1}]\nonumber\\
&=I-A^{p+1}(A^\dagger)^{p+1}.
\end{align}
Taking $p\to\infty$ using $r(A)<1$ yields $I$. Consequently $\sum_m V_m V_m^\dagger=L^\dagger L=I-H_\eta^2$. For columns, $I-A^\dagger A=U^\dagger LL^\dagger U$, and the analogous telescoping identity for $(A^\dagger)^m(I-A^\dagger A)A^m$ gives $\sum_m V_m^\dagger V_m=L^\dagger L$. The direct term in Eq.~\eqref{eq:transferkernel} completes each row and column sum to unity. For $M=M^T$, $U=U^T$ and $(U(KU)^m)^T=(UK)^m U=U(KU)^m$, establishing reciprocity. Equation~\eqref{eq:kernelentropy} follows by exchanging $i$ and $j$ in one copy of the sum and averaging. No spectral continuum approximation enters this argument.

The connection to continuous extended-reservoir relaxation is obtained by taking $\eta_a=e^{-\kappa_a\tau/2}$ with fixed population relaxation rate $\kappa_a>0$. Define $\Gamma=\operatorname{diag}(0,\kappa_{\rm L}I,\kappa_{\rm R}I)$ and $F_0=0\oplus F_{\rm L}\oplus F_{\rm R}$. Expanding $K=I-\tau\Gamma/2+O(\tau^2)$, $U=I-i\tau M+O(\tau^2)$, and $D=\tau\Gamma F_0+O(\tau^2)$ gives
\begin{equation}
\dot C=-i[M,C]-\tfrac12\{\Gamma,C\}+\Gamma F_0.
\label{eq:continuous}
\end{equation}
This is the covariance equation of mode-resolved Markovian relaxation into extended reservoirs. At fixed finite dimension it is the finite-time limit of the cycle map as $\tau\to0$. When the continuous generator is stable, its stationary solution is also the limit of the discrete stationary covariance, by invertibility of the leading continuous Lyapunov operator. This scaling differs from the fixed-$\eta_a<1$ expansion of Sec.~\ref{sec:short}; the singular factor in Eq.~\eqref{eq:gamma} cannot be substituted into the scaling limit term by term. The finite-cycle channel therefore both connects with established continuous-relaxation theory and retains a controllable finite-period regime beyond that limit.

\section{Many-body Gaussian fidelity derivation}\label{app:fidelity}
For a gauge-invariant Gaussian state on $d$ modes with $0<C<I$, diagonalize $C=W\diag(c_1,\ldots,c_d)W^\dagger$. In the corresponding mode basis the density operator is a product of Bernoulli occupation states with probabilities $c_j$. Define $\Gamma(X)=\bigoplus_{n=0}^d\wedge^n X$ on fermionic Fock space, with $\wedge^0 X=1$. The product rule $\Gamma(X)\Gamma(Y)=\Gamma(XY)$ follows sector by sector from the exterior-power representation. For positive $X$, $\Gamma(X)^{1/2}=\Gamma(X^{1/2})$ by diagonalizing $X$. Also $\Tr\Gamma(X)=\prod_j(1+\lambda_j)=\det(I+X)$ for positive $X$, since each Fock occupation contributes the product of its occupied eigenvalues. Thus
\begin{equation}
\hat\rho(C)=\det(I-C)\Gamma(R_C),\quad R_C=C(I-C)^{-1}.
\end{equation}
The eigenvalues of $R_C$ are $c_j/(1-c_j)$, giving the required Bernoulli probabilities and unit trace. Writing $Z_C=\det(I-C)$ and $Z_B=\det(I-B)$,
\begin{align}
\sqrt{\hat\rho(C)}&=\sqrt{Z_C}\Gamma(R_C^{1/2}),\\
\sqrt{\hat\rho(C)}\hat\rho(B)\sqrt{\hat\rho(C)}
 &= Z_C Z_B\Gamma(R_C^{1/2}R_B R_C^{1/2}).
\end{align}
Taking the positive square root, tracing using the determinant identity, and squaring gives Eq.~\eqref{eq:fidexact}. This derivation applies to noncommuting covariances without multiplying their eigenvalues mode by mode. The formula at singular covariances is understood as the continuous limit of full-rank states, not as an inverse of a singular matrix.

For stable numerical evaluation, let $s_j$ be singular values of $R_C^{1/2}R_B^{1/2}$; the eigenvalues of the positive square root in Eq.~\eqref{eq:fidexact} are precisely $s_j$. Then
\begin{equation}
\ln\mathcal F=\sum_j\ln(1-c_j)+\sum_j\ln(1-b_j)+2\sum_j\ln(1+s_j),
\end{equation}
where $b_j$ are the eigenvalues of $B$. This avoids large determinants. The figure uses 141 points on each symmetric cut and 101 on the biased high-retention cut. We independently compared the determinant expression with direct Fock-space evaluation for 32 random full-rank covariance pairs of one to four modes, including noncommuting examples; the largest absolute discrepancy was $3.2\times10^{-15}$. For each point in Fig.~\ref{fig:fidelity} we checked Eqs.~\eqref{eq:fidcontract}, \eqref{eq:fidpresbound}, and \eqref{eq:fidrefbound} within numerical tolerance. The identical-target cut gave $|J_N|<1.3\times10^{-13}$, consistent with the analytic reflection argument. At the near-boundary point $(0.42529,0.9984)$, the symmetric point $(0.9995,0.9995)$, and the singly unrefreshed point $(0.42529,1)$, we independently summed the Lyapunov series by repeated doubling: $X_{j+1}=X_j+B_jX_jB_j^\dagger$ and $B_{j+1}=B_j^2$, starting with $X_0=D$ and $B_0=A$. The residuals were below $2.4\times10^{-15}$; the resulting reference fidelities agreed with the original solver within $2.0\times10^{-13}$ in absolute value. At the singly unrefreshed endpoint the numerical spectral radius is approximately $0.9999999876$, indicating very slow convergence; the covariance difference between solvers was about $2.0\times10^{-8}$ in Frobenius norm. The reported endpoint fidelity is therefore quoted with limited precision. The supplementary reproduction script generates all fidelity data independently of the embedded figure coordinates.

\section{Numerical methods and consistency tests}\label{app:methods}
Constrained local optimization within the parameter square gives the refined estimates in Eq.~\eqref{eq:refined}. At the one-sided nonrefreshing points $(\eta_{\rm L},\eta_{\rm R})=(1,0.25)$ and $(0.25,1)$, the computed currents satisfy $|J_N|<10^{-12}$, consistent with the terminal-blockade theorem. At the refined operating points, independent covariance evaluations of mutual information and relative entropy (using the joint output covariance for the former and the matrix-logarithm definition for the latter) give nonnegative contributions, whose sum agrees numerically with $\sum_a\beta_a(Q_a-\mu_a\Delta N_a)$, as required by Eq.~\eqref{eq:entropyidentity}.\par
For each $N$, construct the real Hermitian one-particle matrix $M$ in its exact finite open-chain eigenmode basis; diagonalize $M$ numerically and set $U=V\diag(e^{-i\epsilon_j\tau})V^\dagger$. Form $K,D$ and solve $C_\fp=A C_\fp A^\dagger+D$ with a dense discrete Lyapunov solver (or equivalently the convergent series \eqref{eq:series}). Extract $C_\fp^-=UC_\fp U^\dagger$, compute Eqs.~\eqref{eq:memory}, \eqref{eq:current} and \eqref{eq:energy}, and verify numerically the stationarity residual, Hermiticity, $0\le C_\fp\le\Id$, and $\Delta N_{\rm L}+\Delta N_{\rm R}=0$.

As an independent check of Eq.~\eqref{eq:transferkernel}, each column of $\mathcal T$ was also computed by solving a separate Lyapunov equation with a unit occupation in the corresponding incoming ancilla mode and zero occupations in all other modes. At the refined current and coherence operating points, the row and column sums differed from unity by less than $5\times10^{-15}$, the maximum reciprocity discrepancy was below $10^{-15}$, and the pairwise entropy production agreed with the direct flux evaluation to within $10^{-13}$. All entries were nonnegative at these points. This calculation does not truncate the sum over earlier injections.

The operating landscapes comprise $41\times41=1681$ parameter pairs for the Gaussian partial-refresh model. Across the grid, the maximum Frobenius norm of the Lyapunov-equation residual was $3.1\times10^{-14}$, the independently reproduced particle-continuity and energy-ledger discrepancies were below $1.2\times10^{-13}$. The smallest eigenvalues of $C_\fp$ and $\Id-C_\fp$ found on this grid were each larger than $1.4\times10^{-5}$; no violation of Eq.~\eqref{eq:memorybound} or the second-law inequality in Eq.~\eqref{eq:secondlaw} was found. These numerical tests concern the specified grid; the positivity and continuity properties also follow analytically. For the algebraic hard-block map, the affine fixed point is obtained with the same one-cycle propagator by solving for the real coordinates $P,\Re x_a,\Im x_a$. Its fermionic admissibility is assessed directly from both Schur-complement inequalities, Eqs.~\eqref{eq:hardlower} and \eqref{eq:hardupper}.

\end{document}